\documentclass[
    reprint,
    superscriptaddress,
    nofootinbib,
    amsmath,
    amssymb,
    aps,
    prd
    ]{revtex4-2}

\usepackage{fix-cm}

\usepackage{graphicx}
\usepackage{dcolumn}
\usepackage{bm}
\usepackage{newtxtext,newtxmath}
\usepackage[T1]{fontenc}
\usepackage{amsmath}
\usepackage{glossaries}
\glsdisablehyper
\usepackage{xspace}
\usepackage{multirow}
\usepackage{verbatim}
\usepackage{bm}
\usepackage{hyperref}
\usepackage[dvipsnames]{xcolor}

\DeclareRobustCommand{\VAN}[3]{#2}
\let\VANthebibliography\thebibliography
\def\thebibliography{\DeclareRobustCommand{\VAN}[3]{##3}\VANthebibliography}

\newcommand{\shape}{{\bm \Theta_{\rm s}}}
\newcommand{\evolution}{{\bm \Theta_{\rm e}}}
\newcommand{\shapestar}{{\bm \Theta_{{\rm s},\star}}}

\newcommand{\shapeplanck}{{\bm \Theta_{\rm s}^{\, \Planck}}}
\newcommand{\evolutionplanck}{{\bm \Theta_{\rm e}^{\,\Planck}}}
\newcommand{\evolutionfixed}{{\bm \Theta_{\rm e}^{\rm fixed}}}
\newcommand{\nuis}{{\bm \Theta_{\rm nuis}}}
\newcommand{\fiducial}{{\bm \Theta^{\rm fid}}}
\newcommand{\omm}{\omega_{\rm m}}
\newcommand{\omc}{\omega_{\rm c}}
\newcommand{\omcstar}{\omega_{{\rm c},\,\star}}
\newcommand{\omctilde}{\tilde{\omega}_{\rm c}}
\newcommand{\omb}{\omega_{\rm b}}
\newcommand{\omg}{\omega_{\gamma}}
\newcommand{\omde}{\omega_{\rm DE}}
\newcommand{\omnu}{\omega_\nu}
\newcommand{\omnustar}{\omega_{\nu,\,\star}}

\newcommand{\ommo}{\omega_{\rm m,0}}

\newcommand{\HO}{H_0}
\newcommand{\wo}{w_0}
\newcommand{\wa}{w_a}
\newcommand{\Mnu}{M_{\nu}}
\newcommand{\Nnu}{N_{\nu}}
\newcommand{\ns}{n_{\rm s}}
\newcommand{\rns}{\frac{{\rm d}\,\ns}{{\rm d}\log k}}
\newcommand{\As}{A_{\rm s}}

\newcommand{\omk}{\omega_{\rm K}}
\newcommand{\sigtwelve}{\sigma_{12}}
\newcommand{\sigtwelvestar}{\sigma_{12,\star}}
\newcommand{\kmsMpc}{{\rm km\,s^{-1}\,{Mpc}^{-1}}}
\newcommand{\Plin}{P_{\rm lin}}
\newcommand{\Pstar}{P_\star}
\newcommand{\Pcb}{P_{\rm cb}}

\newcommand{\Pcbnu}{P_{\rm cb\nu}}
\newcommand{\deltacb}{\delta_{\rm cb}}
\newcommand{\deltanu}{\delta_\nu}
\newcommand{\Tcb}{T_{\rm cb}}
\newcommand{\Tnu}{T_\nu}
\newcommand{\Tcbnu}{T_{\rm cb\nu}}
\newcommand{\fcb}{f_{\rm cb}}
\newcommand{\fnu}{f_\nu}
\newcommand{\Mpc}{{\rm Mpc}}
\newcommand{\Mpch}{h^{-1}\,{\rm Mpc}}
\newcommand{\kMpc}{{\rm Mpc}^{-1}}
\newcommand{\kMpch}{h\,{\rm Mpc}^{-1}}
\newcommand{\lams}{\lambda_{\rm s}}
\newcommand{\cnu}{c_\nu}
\newcommand{\eV}{{\rm eV}}
\newcommand{\comet}{{\tt COMET}\xspace}

\newcommand{\Pnw}{P_{\rm nw}}
\newcommand{\Pw}{P_{\rm w}}
\newcommand{\Peh}{P_{\rm EH}}
\newcommand{\logten}{\log_{10}}
\newcommand{\lcdm}{$\Lambda$CDM\xspace}
\newcommand{\wowacdm}{$\wo\wa$CDM\xspace}
\newcommand{\bone}{b_1}
\newcommand{\btwo}{b_2}
\newcommand{\gtwo}{\gamma_2}
\newcommand{\gtwoone}{\gamma_{21}}
\newcommand{\czero}{c_0}
\newcommand{\ctwo}{c_2}
\newcommand{\cfour}{c_4}
\newcommand{\cnlo}{c_{\rm nlo}}
\newcommand{\npzero}{N^P_0}
\newcommand{\nptwozero}{N^P_{20}}
\newcommand{\nptwotwo}{N^P_{22}}
\newcommand{\bnablatwo}{b_{\nabla^2}}
\newcommand{\sqdeg}{{\rm sqdeg}}
\newcommand{\Euclid}{{\it Euclid}\xspace}
\newcommand{\Planck}{{\it Planck}\xspace}

\newcommand{\vtwo}{{\tt v2}\xspace}
\newcommand{\Pell}{P_{\ell}}
\newcommand{\kv}{{\bm k}}
\newcommand{\qv}{{\bm q}}
\newcommand{\sigv}{\sigma_v}
\newcommand{\qperp}{q_\perp}
\newcommand{\qpara}{q_\parallel}
\newcommand{\kperp}{k_\perp}
\newcommand{\kpara}{k_\parallel}
\newcommand{\DM}{D_M}
\newcommand{\Dh}{D_H}
\newcommand{\legell}{{\cal L}_\ell}
\newcommand{\Winfty}{W_\infty}

\newcommand{\kmin}{k_{\rm min}}
\newcommand{\kmax}{k_{\rm max}}
\newcommand{\gpy}{{\tt GPy}\xspace}
\newcommand{\deltag}{\delta_{\rm g}}
\newcommand{\Pgg}{P_{\rm gg}}
\newcommand{\xv}{{\bm x}}
\newcommand{\yv}{{\bm y}}
\newcommand{\sv}{{\bm s}}
\newcommand{\rv}{{\bm r}}
\newcommand{\lv}{{\bm l}}
\newcommand{\vvv}{{\bm v}}
\newcommand{\uv}{{\bm u}}

\newcommand{\numpy}{{\tt numpy}\xspace}
\newcommand{\vpara}{v_\parallel}
\newcommand{\upara}{u_\parallel}
\newcommand{\epara}{{\bm e}_\parallel}
\newcommand{\deltas}{\delta^{\,s}}
\newcommand{\nablapara}{\nabla_\parallel}
\newcommand{\avir}{a_{\rm vir}}
\newcommand{\Pctr}{P_{\rm ctr}}
\newcommand{\Ps}{P^{\,s}}
\newcommand{\Pggs}{P_{\rm gg}^{\,s}}
\newcommand{\Pctrlo}{\Pctr^{\,\rm LO}}
\newcommand{\Pctrnlo}{\Pctr^{\,\rm NLO}}
\newcommand{\PggEFT}{P_{\rm gg}^{\,\rm EFT}}
\newcommand{\Pggtree}{P_{\rm gg}^{\,\rm tree}}
\newcommand{\Pggoneloop}{P_{\rm gg}^{\,\rm 1\mbox{-}loop}}
\newcommand{\Pggctr}{P_{\rm gg}^{\,\rm ctr}}
\newcommand{\Pggnoise}{P_{\rm gg}^{\,\rm stoch}}
\newcommand{\PggVDG}{P_{\rm gg}^{\,\rm VDG_\infty}}
\newcommand{\Pdd}{P_{\delta\delta}}
\newcommand{\Pdt}{P_{\delta\theta}}
\newcommand{\Ptt}{P_{\theta\theta}}
\newcommand{\Neff}{N_{\rm eff}}
\newcommand{\Nur}{N_{\rm ur}}

\newacronym{cmb}{CMB}{cosmic microwave background}
\newacronym{eft}{EFT}{effective field theory of large-scale structure}
\newacronym{vdg}{VDG$_\infty$}{velocity difference generating function}
\newacronym{bao}{BAO}{baryon acoustic oscillations}
\newacronym{rpt}{RPT}{renormalised perturbation theory}
\newacronym{dst}{DST}{\emph{discrete sine transform}}
\newacronym{ir}{IR}{infrared}
\newacronym{cdm}{CDM}{cold dark matter}
\newacronym{eds}{EdS}{Einstein-de Sitter}
\newacronym{rsd}{RSD}{redshift-space distortions}
\newacronym{apdist}{AP}{Alcock-Paczynski}
\newacronym{gp}{GP}{Gaussian process}
\newacronym{rbf}{RBF}{radial basis function}
\newacronym{scikit}{SL}{\texttt{scikit-learn}}
\newacronym{lhc}{LHC}{Latin hypercube}
\newacronym{lss}{LSS}{large-scale structure}
\newacronym{fog}{FoG}{finger-of-God}
\newacronym{hod}{HOD}{halo occupation distribution}
\newacronym{pt}{PT}{perturbation theory}

\newcommand{\eft}{\gls{eft}\xspace}
\newcommand{\vdg}{\gls{vdg}\xspace}
\newcommand{\bao}{\gls{bao}\xspace}

\newcommand{\dst}{\gls{dst}\xspace}
\newcommand{\ir}{\gls{ir}\xspace}
\newcommand{\cdm}{\gls{cdm}\xspace}
\newcommand{\eds}{\gls{eds}\xspace}
\newcommand{\rsd}{\gls{rsd}\xspace}
\newcommand{\apdist}{\gls{apdist}\xspace}
\newcommand{\gp}{\gls{gp}\xspace}
\newcommand{\rbf}{\gls{rbf}\xspace}
\newcommand{\scikit}{\gls{scikit}\xspace}
\newcommand{\lhc}{\gls{lhc}\xspace}
\newcommand{\lss}{\gls{lss}\xspace}
\newcommand{\cmb}{\gls{cmb}\xspace}
\newcommand{\fog}{\gls{fog}\xspace}
\newcommand{\hod}{\gls{hod}\xspace}

\newcommand{\python}{\texttt{Python}\xspace}
\newcommand{\camb}{\texttt{CAMB}\xspace}
\newcommand{\classy}{\texttt{CLASS}\xspace}

\newcommand\aap{A\&A}                
\newcommand\apjl{ApJ}                
\newcommand\jcap{J.~Cosmology Astropart. Phys.} 
\newcommand\mnras{MNRAS}             
\newcommand\physrep{Phys.~Rep.}      

\begin{document}

\preprint{APS/123-QED}

\title{Extending evolution mapping to massive neutrinos with {\tt COMET}}

\author{Andrea Pezzotta}
\email{andrea.pezzotta@inaf.it}
\affiliation{Max Planck Institute for Extraterrestrial Physics, Giessenbachstr. 1, 85748 Garching, Germany}
\affiliation{INAF - Osservatorio Astronomico di Brera, via Emilio Bianchi 46, 23807
Merate, Italy}

\author{Alexander Eggemeier}
\affiliation{Universit\"at Bonn, Argelander-Institut f\"ur Astronomie, Auf dem H\"ugel 71, 53121 Bonn, Germany}

\author{Giosuè Gambardella}
\affiliation{Institute of Space Sciences (ICE, CSIC), Campus UAB, Carrer de Can Magrans, s/n, 08193 Barcelona, Spain}
\affiliation{Institut d'Estudis Espacials de Catalunya (IEEC), Carrer Gran Capit\'a 2-4, 08034 Barcelona, Spain}

\author{Lukas Finkbeiner}
\affiliation{Max Planck Institute for Extraterrestrial Physics, Giessenbachstr. 1, 85748 Garching, Germany}

\author{Ariel G. Sánchez}
\affiliation{Max Planck Institute for Extraterrestrial Physics, Giessenbachstr. 1, 85748 Garching, Germany}

\author{Benjamin Camacho Quevedo}
\affiliation{Institute of Space Sciences (ICE, CSIC), Campus UAB, Carrer de Can Magrans, s/n, 08193 Barcelona, Spain}
\affiliation{Institut d'Estudis Espacials de Catalunya (IEEC), Carrer Gran Capit\'a 2-4, 08034 Barcelona, Spain}
\affiliation{Institute for Fundamental Physics of the Universe, Via Beirut 2, 34151 Trieste, Italy}
\affiliation{Scuola Internazionale di Studi Superiori Avanzati, via Bonomea 265, 34136 Trieste, Italy}
\affiliation{INAF - Osservatorio Astronomico di Trieste, via Tiepolo 11, 34143 Trieste, Italy}

\author{Martin Crocce}
\affiliation{Institute of Space Sciences (ICE, CSIC), Campus UAB, Carrer de Can Magrans, s/n, 08193 Barcelona, Spain}
\affiliation{Institut d'Estudis Espacials de Catalunya (IEEC), Carrer Gran Capit\'a 2-4, 08034 Barcelona, Spain}

\author{Nanoom Lee}
\affiliation{Center for Cosmology and Particle Physics, Department of Physics, New York University, NY 10003, New York, USA}
\affiliation{William H. Miller III Department of Physics \& Astronomy, Johns Hopkins University, Baltimore, MD 21218, USA}

\author{Gabriele Parimbelli}
\affiliation{Institute of Space Sciences (ICE, CSIC), Campus UAB, Carrer de Can Magrans, s/n, 08193 Barcelona, Spain}
\affiliation{Scuola Internazionale di Studi Superiori Avanzati, via Bonomea 265, 34136 Trieste, Italy}

\author{Román Scoccimarro}
\affiliation{Center for Cosmology and Particle Physics, Department of Physics, New York University, NY 10003, New York, USA}

\date{\today}


\begin{abstract}
We introduce an extension of the evolution mapping framework to cosmological models that include massive neutrinos. The original evolution mapping framework exploits a degeneracy in the linear matter power spectrum when expressed in Mpc units, which compresses its dependence on cosmological parameters into those that affect its shape and a single extra parameter $\sigtwelve$, defined as the RMS linear variance in spheres of radius 12 Mpc. We show that by promoting the scalar amplitude of fluctuations, $\As$, to a shape parameter, we can additionally describe the suppression due to massive neutrinos at any redshift to sub-0.01\% accuracy across a wide range of masses and for different numbers of mass eigenstates. This methodology has been integrated into the public \comet package, enhancing its ability to emulate predictions of state-of-the-art perturbative models for galaxy clustering, such as the effective field theory (EFT) model. Additionally, the updated software now accommodates a broader cosmological parameter space for the emulator, enables the simultaneous generation of multiple predictions to reduce computation time, and incorporates analytic marginalization over nuisance parameters to expedite posterior estimation. Finally, we explore the impact of different infrared resummation techniques on galaxy power spectrum multipoles, demonstrating that any discrepancies can be mitigated by EFT counterterms without impacting the cosmological parameters.
\end{abstract}


\maketitle

\section{Introduction}

The advent of Stage-IV experiments represents a remarkable milestone in the field of observational cosmology and the study of the \lss of the Universe. Observational projects such as the Dark Energy Spectroscopic Instrument \citep[DESI;][]{DESI2016}, \Euclid \citep{EUCLID2011, MelAbdAce2024}, the Nancy Grace Roman Space Telescope \citep[Roman;][]{ROMAN2015}, and the The Rubin Observatory Legacy Survey of Space and Time \citep[LSST;][]{LSST2008}, will collect galaxy samples comprising tens of millions of sources. These datasets will thus enable the reconstruction of a 3D map of the observable Universe to an unprecedented level of detail.


Harnessing the full potential of these datasets requires the development of efficient tools capable of jointly analyzing multiple cosmological probes. By integrating different observables within a unified framework, it becomes possible to break parameter degeneracies and mitigate prior volume effects in parameter estimation \citep{HadWolAzz2023, CarMorPou2023, SimZhaPou2023}, ultimately enhancing the robustness of cosmological constraints. At the same time, the need for fast and accurate methods to evaluate theoretical models is becoming increasingly pressing, particularly at non-linear scales, where computational costs would otherwise be prohibitive.

Over the last decade, the cosmological community has made significant progress in the development of emulators -- computational tools designed to provide smart multi-dimensional interpolations of various statistics across the cosmological parameter space. The primary advantage of emulators is their ability to replicate different clustering statistics as measured from high-precision N-body simulations, while significantly reducing computational time. This approach has been widely applied to the matter power spectrum \citep{LawHeiWhi2010, ZhaTinBac2019, NisTakTak2019, KnaStaPot2021, AngZenCon2021} , as well as to other statistics beyond dark matter clustering, such as baryonic corrections \citep{AriAngCon2021} and the large-scale distribution of galaxies. For the latter, different strategies exist. One approach bypasses the need to emulate galaxy bias parameters and directly emulates the individual blocks of the standard one-loop perturbative expansion of the galaxy power spectrum \citep{ZenAngPel2023, PelAngZen2023}. Alternatively, it is possible to train emulators to reproduce the final galaxy power spectrum on N-body simulations that reproduce different \hod galaxy samples, either incorporating the \hod parameters in the parameter space of the emulator \citep{KwaHeiHab2015, ZhaTinBac2019} or analytically modelling the galaxy-halo distribution a posteriori \citep{KobNisTak2020}.

An alternative approach focuses on constructing emulators for theory codes that rely on a perturbative expansion of the density and velocity fields \citep{ChuIvaPhi2020, AmiSenZha2021, EggCamPez2022, McCKoyBeu2023, MauCheWhi2024, LinMorRad2024, BonAmiBel2025}. This method preserves the accuracy of perturbative methods on mildly non-linear scales while accelerating the computation of non-linear models by several orders of magnitude. However, the development of accurate emulators faces significant challenges due to the growing parameter space that must be explored when considering extensions to the standard cosmological model. This issue is particularly important in light of the ambitious goals of Stage-IV surveys, which aim to place percent-level constraints on key cosmological parameters, such as the equation of state of dark energy, the curvature of the Universe, and the total mass of neutrinos. 

In this context, evolution mapping \citep{SanRuiJar2022} offers an innovative way of addressing the high dimensionality of the cosmological parameter spaces. This approach captures the complex interplay of various cosmological parameters, referred to as evolution parameters, by capturing their combined effect on the matter density field through a single derived quantity, $\sigtwelve$ \citep{Sanchez2020}. This parameter represents the variance of the linear density field on a fixed scale in units of $\Mpc$, as opposed to the more conventional choice of using $\Mpch$ units. By parametrising the evolution of the matter density field in terms of $\sigtwelve$, one can show that different combinations of evolution parameters produce equivalent density fields, provided they share the same value of $\sigtwelve$. Furthermore, for cosmologies with the same power spectrum shape, the evolution of the density field can be consistently rescaled across different sets of evolution parameters by expressing it in terms of $\sigtwelve$.

The degeneracy experienced by evolution parameters is exact at the linear level, but \cite{SanRuiJar2022} demonstrated that this relation holds approximately even in the highly non-linear regime, on scales around $k\sim 1\,\kMpc$ and for redshifts approaching $z=0$. More recently, \cite{EspSanBel2024} extended evolution mapping by showing that it can also effectively describe the time evolution of the velocity field, quantified through the auto velocity-velocity and cross density-velocity power spectra ($P_{\theta\theta}$ and $P_{\delta\theta}$), obtaining consistent accuracy levels. In addition, \cite{EggCamPez2022} and \cite{EggLeeSco2024} applied the evolution mapping framework to construct \comet, a theory emulator for \rsd models that also includes the \eft \citep[e.g.][]{CarHerSen2012, PerSenJen2016, IvaSimZal2020} and the recently developed framework based on the \vdg, effectively propagating the degeneracy to also describe the clustering of biased tracers in redshift space, and reducing the input parameter space over which the training has to be performed. The emulator, publicly available through PyPI\,\footnote{\url{https://pypi.org/project/comet-emu}} and GitLab\,\footnote{\url{https://gitlab.com/aegge/comet-emu}}, has already been exploited in the context of prelaunch analyses within the Euclid Consortium for efficiently sampling posterior distributions in galaxy clustering likelihood analyses, as shown in recent works \citep[e.g.][]{PezMorZen2024}.

However, to meet the stringent precision requirements of Stage-IV surveys, a critical limitation of the evolution mapping framework must be addressed: the proper incorporation of massive neutrinos. Unlike cold dark matter or baryons, neutrinos possess significant thermal velocities, which allow them to free-stream out of overdense regions, suppressing the growth of structure on scales below their free-streaming length \citep[see e.g.][for a comprehensive review on the subject]{LesPas2006}. This suppression introduces a scale-dependent growth factor, fundamentally complicating the rescaling of the amplitude of density fluctuations across redshift. Moreover, the magnitude of this effect depends on the total neutrino mass and the number of massive species, adding further complexity to the problem. As such, the standard evolution mapping framework, which assumes a scale-independent growth factor, is inadequate for cosmologies including massive neutrinos. Extending the framework to incorporate these effects is not just an improvement but an essential step toward ensuring its applicability to the broader parameter spaces probed by next-generation cosmological surveys.

This work proposes a novel mapping applicable to the suppression factor induced by massive neutrinos on the matter power spectrum, showing that it relies on the relative growth factor across cosmologies compared to initial density fluctuation amplitudes. With this method, cosmologies can still be mapped using $\sigtwelve$ as a time variable, maintaining consistency with the original evolution mapping approach. We implemented this extension in \comet, achieving evaluation speeds for non-linear models with massive neutrinos that are three orders of magnitude faster than direct theory code evaluations (from ${\cal O}(10\,{\rm s})$ to ${\cal O}(10\,{\rm ms})$ on a single CPU). The new version software, labelled \vtwo, has been trained to reproduce theory predictions for the two aforementioned \rsd frameworks, \eft and \vdg, whose relative performances have been recently benchmarked in \cite{EggLeeSco2024}. Additional features include a broadening of the input cosmological parameter space used to train the emulator, the simultaneous generation of multiple theory predictions for reduced evaluation time, the analytical marginalization of linear parameters that facilitates a faster posterior estimation, and finally the use of a more robust algorithm to account for the impact of infrared modes on the amplitude of the baryon acoustic peaks in the galaxy power spectrum. With this article, we officially make this second version of \comet publicly available. 

This paper is structured in the following way. In Sect. \ref{sec:evolution_mapping} we first revisit the original formulation of evolution mapping and then proceed to extend it to cosmologies including massive neutrinos. In Sect. \ref{sec:models} we provide an overview of the \eft and \vdg models, focusing on the different treatment of \rsd at separations approaching the nonlinear scale. In Sect. \ref{sec:emulator_design} we present the design and extra features of the new version of \comet, while in Sect. \ref{sec:ValPerf} we discuss the validation and performance of the code and examine how different techniques that resum  infrared modes to account for the damping of \bao affect the recovery of cosmological parameters. Finally, in Sect. \ref{sec:conclusions} we give our conclusions.

\section{Evolution mapping}
\label{sec:evolution_mapping}

In the following subsections, we summarize the baseline formulation of evolution mapping in the context of a massive-neutrino-free cosmology, focusing on the mapping between different cosmologies at the level of the linear power spectrum \citep{SanRuiJar2022}. We then proceed to describe the minimal setup required to extend this framework to include the effects of massive neutrinos, which is one of the main results of this paper.

\subsection{Original formulation}
\label{sec:original_formulation}

The main aspect behind evolution mapping relies on the distinction between two different subsets of cosmological parameters, \emph{shape} and \emph{evolution}, respectively. The shape parameters,
\begin{equation}
    \shape = \left\{\omb,\, \omc,\, \omg,\, \ns,\, \rns,\, \ldots\right\},
\end{equation}
directly impact the shape of the matter transfer function $T(k)$ and the post-inflationary primordial power spectrum $P_{\cal R}(k)$. They are comprised of the physical density parameters, $\omega_{\rm X}\equiv\Omega_{\rm X}h^2$, of different components, such as baryons ($\omb$), cold dark matter ($\omc$), and radiation ($\omg$), in addition to the scalar primordial index ($\ns$), its running with the wave mode $k$, and potentially other parameters from extensions to the standard cosmological model. On the other side, the evolution parameters,
\begin{equation}
    \evolution = \left\{\omk, \, \omde, \, \wo, \, \wa, \, \As, \, \ldots \right\},
\end{equation}
only affect the overall amplitude of the matter power spectrum at any later (than the matter-radiation equality) epoch. They include the physical density parameters of curvature ($\omk$) and dark energy ($\omde$), together with the parameters describing its equation of state ($\wo$, $\wa$),\footnote{This is based on the well-known CLP parametrisation \citep{ChePol2001, Linder2003}, for which the time evolution of the dark energy equation of state is described as $$w(z)=\wo+\wa\frac{z}{1+z}\,.$$ However, any other standard parametrization may be added as a further set of evolution parameters.} the scalar primordial amplitude ($\As$), and similarly other parameters that do not modify the shape of the power spectrum. In this context, the Hubble expansion rate ($h\equiv\HO/(100\,\kmsMpc)$) is a mixture of shape and evolution parameters, since from the first Friedmann equation it follows that
\begin{equation}
    h^2=\sum_i\omega_i\,,
    \label{eq:sum_densities}
\end{equation}
 with the sum running over all the components of the Universe. However, for fixed shape parameters, any change in $h$ effectively leads to a modification in the total dark energy density $\omde$. Likewise, the fractional density parameters ($\Omega_i\equiv\omega_i/h^2$) are also a combination of shape and evolution parameters.

For fixed shape parameters, the effect of changing the evolution parameters can then be described using a single quantity, or, in different words, evolution parameters are degenerate with each other in terms of the actual amplitude of linear density fluctuations.\footnote{This statement is valid only over a restricted range of separations, since on large-enough scales strong deviations from a cosmological constant ($w=-1$) leads to ${\cal O}(1\%)$ broadband differences at $k\lesssim 0.01\,\kMpc$, in the same way as curvature modifies the amplitude of the linear power spectrum on superhorizon scales. However, for the range of scales that are typically employed in galaxy clustering analyses ($k>0.01\,\kMpc$), this does not represent a limitation to the applicability of evolution mapping.} Traditionally, the amplitude of the linear power spectrum is described in terms of the rms fluctuations of the density field within some fixed physical scale $R$, as
\begin{equation}
    \sigma^2(R) = \frac{1}{2\pi^2} \int k^2 \, \Plin(k) \, W^2_{\rm TH}(kR) \, {\rm d}k \,,
\end{equation}
where
\begin{equation}
    W_{\rm TH}(kR) = 3\left[\frac{\sin(kR)}{(kR)^3} - \frac{\cos(kR)}{(kR)^2}\right]
\end{equation}
is the Fourier transform of a top-hat window function of size $R$. In this context, selecting a reference value of $R$ in units of $\Mpc$ is crucial, rather than the conventional choice of $R=8\,\Mpch$. This is necessary because varying the reference value of $h$ would otherwise affect $k$ and $P(k)$ differently and thus break the degeneracy of the evolution parameters \citep{Sanchez2020}. For a \Planck-like value of $h\approx0.67$, the previous scale translates approximately to $12 \,\Mpc$, making $\sigtwelve\equiv\sigma(12\,\Mpc)$ a reasonable choice to describe the degeneracy experienced by evolution parameters.

With these premises, it is then possible to derive an expression for the linear matter power spectrum in terms of the shape and evolution parameters, as
\begin{equation}
    \Plin(k\,|\,\shape,\,\evolution,\,z) = \Plin\left(k\,|\,\shape,\,\sigtwelve(\shape,\,\evolution,\,z)\right)\,.
    \label{eq:orig_evmap}
\end{equation}
This means that, once expressed in $\Mpc$ units, the dependence of the power spectrum on the whole set of cosmological parameters can be reduced to the dependence on the shape parameters and on $\sigtwelve$, which effectively acts as a proxy for the amplitude of the power spectrum. The dependence on the redshift $z$ can also be absorbed into $\sigtwelve$, since time evolution only induces an overall constant increase of power via the growth factor $D(z)$.

\begin{figure}
    \includegraphics[width=\columnwidth]{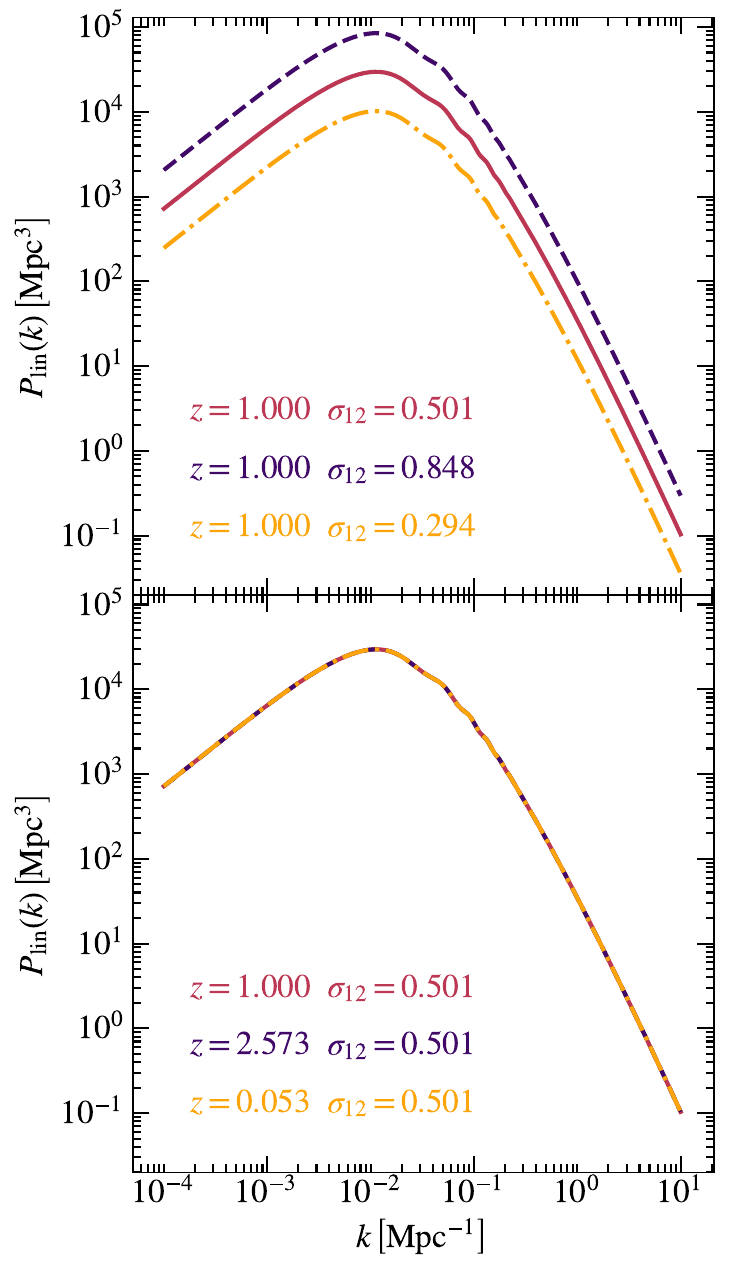}
    \caption{Evolution of the linear power spectrum for cosmologies with the same set of shape parameters $\shape$. In the top panel, the power spectra are evaluated at a fixed reference redshift ($z=1$), showing differences in amplitude. In the bottom panel, they are evaluated at redshifts chosen to match the same value of $\sigtwelve$, resulting in identical amplitudes.}
    \label{fig:evomap_nonu_linear}
\end{figure}

This behavior is illustrated in Fig. \ref{fig:evomap_nonu_linear}, where we show the time evolution of the linear power spectrum for three different cosmologies sharing the same set of shape parameters but with varying evolution parameters. In all cases, we obtain predictions using the public Boltzmann solver \camb,\footnote{\url{https://github.com/cmbant/CAMB}} and we have verified that these predictions are consistent with those obtained from the \classy\,\footnote{\url{https://github.com/lesgourg/class_public}} Boltzmann solver at the level of ${\cal O}(0.01\mbox{-}0.1\%)$. Considering configurations with the same redshift ($z=1$ in this case) leads to different amplitudes for the three spectra, a consequence of the different growth factors $D(z)$ of the three cosmologies. However, by considering individual redshifts resulting in the same value of $\sigtwelve$ it is possible to obtain consistent amplitudes across the entire range of separations considered, as can be observed in the bottom panel.

While the degeneracy between evolution parameters is perfect at linear level, \cite{SanRuiJar2022} showed that this can be also propagated to the non-linear regime, with small residuals appearing at the level of the one-halo term that are sourced by different structure growth histories.\footnote{The residual discrepancy mentioned above is, in any case, irrelevant within the framework of perturbative models of galaxy clustering under the \eds approximation. This is because any remaining dependence on the chosen cosmology vanishes, and the model becomes fully determined by $\Plin$, which can be perfectly mapped with evolution mapping.} This can be corrected by properly accounting for the different structure formation histories, quantified in terms of the suppression factor  $g(z)=(1+z)\,D(z)$, leading to ${\cal O}(0.1\%)$ accuracies up to $k=1.5\,\kMpc$ for a broad range of $\sigtwelve$ values.\footnote{The accuracy of the mapping slightly decreases for higher values of $\sigtwelve$, or alternatively at lower redshifts, due to the stronger influence of non-linear evolution in this regime. Nevertheless, even when considering snapshots with $\sigtwelve\sim0.83$, different cosmologies are in agreement at better than $1\%$, and most of the discrepancies are concentrated in the strongly non-linear regime ($k\gtrsim1\,\kMpc$).} More recently, \cite{EspSanBel2024} extended the validity of this formalism from the density $\delta$ to the velocity field $\theta$, showing that also the velocity power spectra -- auto $\Ptt$ and cross $\Pdt$ -- can be mapped from one cosmology to the other in terms of $\sigtwelve$ with similar levels of accuracy.

\subsection{Extension to massive neutrinos}

The framework described in the previous section can be used to characterize the evolution of the density and velocity fields across vastly different cosmologies. However, a key requirement for its validity is the absence of a massive neutrino component contributing to the total energy budget of the Universe. The fundamental issue is that, for a fixed number of effective neutrino species $\Neff$,  massive neutrinos, or more specifically, the total neutrino mass $\Mnu$, do not fit neatly into either the shape or evolution parameter categories but instead act as a mixture of the two. This is because massive neutrinos modify the shape of the linear power spectrum through a scale-dependent growth factor $D(z,k)$, where time dependence can no longer be separated from spatial dependence due to the damping effect caused by neutrino free streaming, which evolves over time.

\begin{figure}
    \includegraphics[width=\columnwidth]{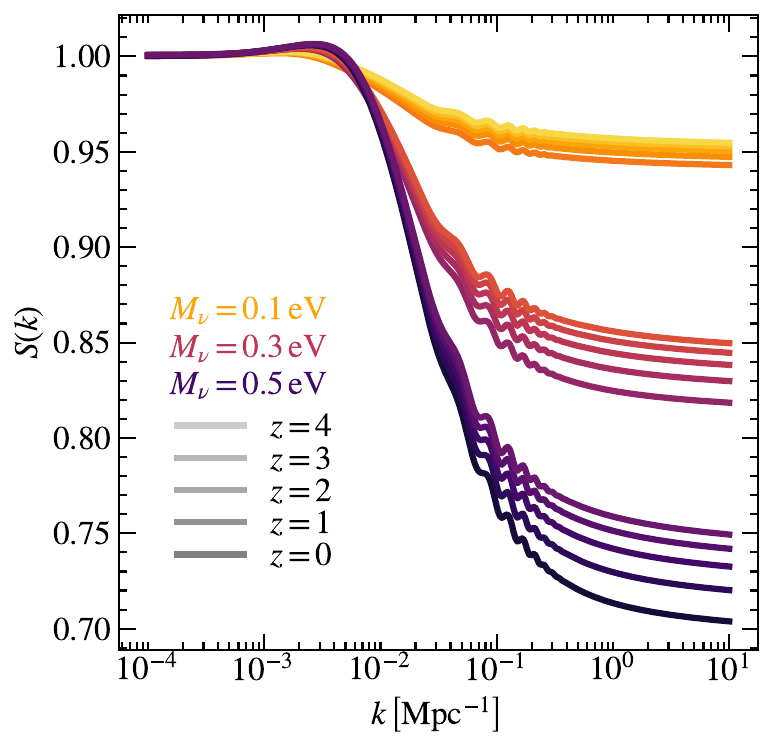}
    \caption{Evolution of the neutrino suppression factor as a function of the different total neutrino mass $M_\nu$, redshift $z$, and wave number $k$. Different variations of the same colour correspond to different values of $z$, with the darkest shade corresponding to the case at $z=0$, for which the neutrino damping is strongest.}
    \label{fig:supp_factor}
\end{figure}

This can be easily observed in Fig. \ref{fig:supp_factor}, where we plot the time evolution of the suppression factor,
\begin{equation}
    S(k)\equiv \frac{P(k)}{\Pstar(k)}\,,
    \label{eq:suppression_factor}
\end{equation}
for three different values of the total neutrino mass $\Mnu$. In this case as well, predictions have been obtained using \camb, but we have explicitly verified that consistent results (within 0.1\%) can be obtained with \classy, provided the default fluid approximation for massive neutrinos is disabled.\footnote{Specifically, in this comparison we adopted \texttt{ncdm\_fluid\_approximation = 3} and \texttt{l\_max\_ncdm = 40}, which correspond to a full Boltzmann hierarchy integration for massive neutrinos up to multipole $\ell=40$, without invoking any fluid approximation.}

In Eq.~(\ref{eq:suppression_factor}), we use $P$ to denote the linear power spectrum of a cosmology that includes massive neutrinos, and $\Pstar$ to represent the corresponding spectrum of an equivalent cosmology with the same total dark matter density, but composed entirely of cold dark matter. The latter is defined by $\shapestar$, where $\omcstar=\omc+\omnu$ and $\omnustar=0$. In this comparison, the cosmology with massive neutrinos assumes one massive species and two massless ones, and adopts an effective number of relativistic species $\Nur=2.029$, so that the total energy density in relativistic species corresponds to an effective neutrino number $\Neff=3.044$. Two main effects can be seen in Fig. \ref{fig:supp_factor}. First, since the total matter density is conserved, the presence of massive neutrinos delays the time of matter-radiation equality, as indicated by the small excess of power at $k\sim0.005\,\kMpc$. Secondly, the growth of linear perturbations is suppressed on scales much smaller than the free-streaming length of any massive neutrino species \citep{LesPas2006}. For an individual species with mass  $m_\nu$, the comoving free-streaming scale can be approximated as
\begin{equation}
        \lams(z) = 2\pi\sqrt{\frac{2}{3}}\frac{\cnu(z)}{H(z)} \simeq
         7.7\sqrt{\frac{\omm(z)}{\ommo(1+z)}}\left(\frac{1\eV}{m_\nu}\right)\Mpc\,,
\end{equation}
where $\omm(z)$ and $\ommo$ are the physical matter density parameter at the considered redshift and at present time, respectively, $H(z)$ is the Hubble expansion rate, and $\cnu$ is the neutrino thermal velocity. As time passes, the amplitude of the damping gets larger, as massive neutrinos suppress the growth of density fluctuations below the free-streaming scale. 
This implies that the evolution mapping principle breaks down in this scenario, as redshift can no longer be used to track the evolution of the density field in a parameter-independent way. This is especially noticeable when exploring cosmologies with a sufficiently large value of $\Mnu$, like $\Mnu\sim0.5$-$1 \ \eV$, which have not been completely discarded by galaxy clustering constraints from Stage-III experiments \citep{SemSanPez2023, MorTseCar2023}.

To address the residual effect at scales smaller than the neutrino free-streaming scale within the context of evolution mapping, we can rearrange Eq. (\ref{eq:suppression_factor}) and express the linear power spectrum for a generic cosmology (including massive neutrinos) as the product of the neutrino-free power spectrum and the suppression factor,
\begin{equation}
    P(k) = \Pstar(k)\,S(k)\,.
\end{equation}
As discussed in Sect. \ref{sec:original_formulation}, $\Pstar$ can be mapped between different cosmologies solely based on the shape parameters $\shapestar$ and the total amplitude $\sigtwelvestar$. The key question is then whether it is feasible to develop a new mapping for the suppression factor $S$ that would allow evolution mapping to effectively apply to $P$ as well.

\begin{figure*}
    \includegraphics[width=2.1\columnwidth]{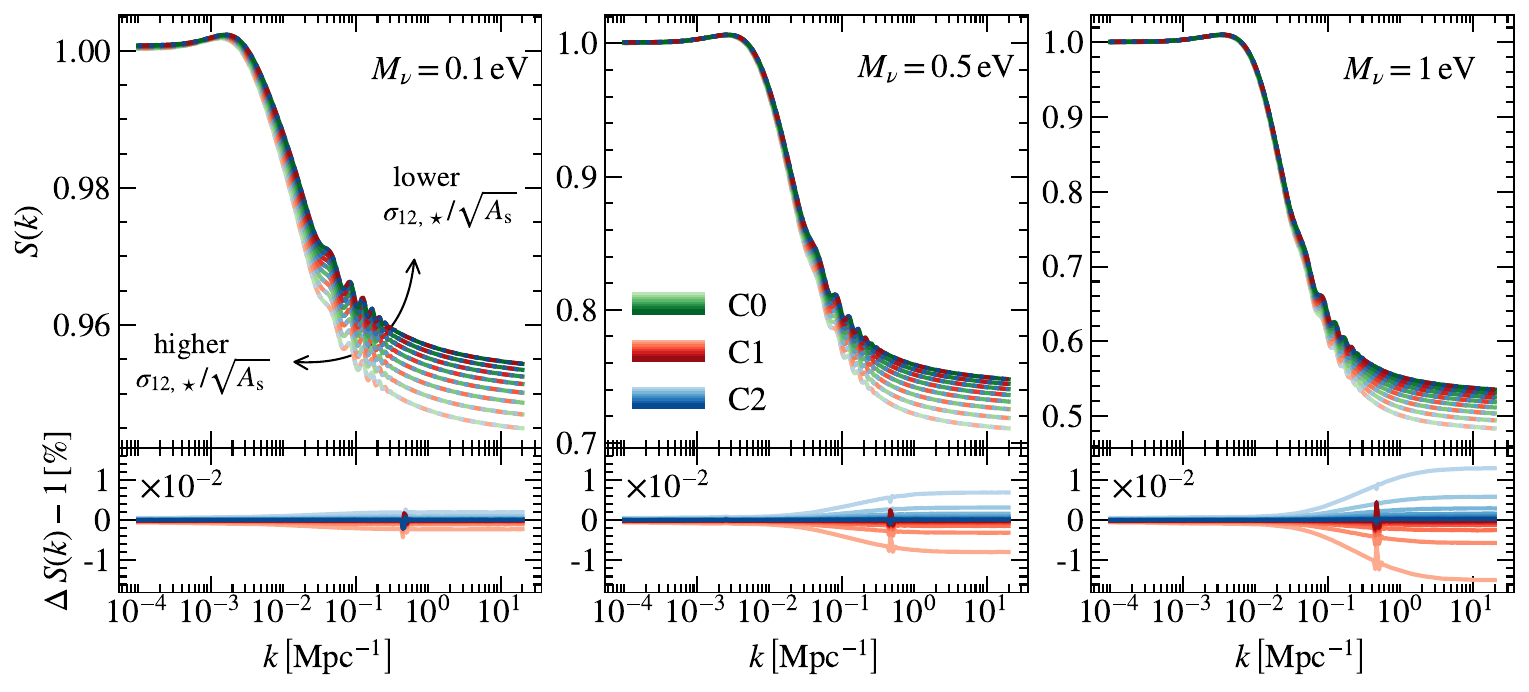}
    \caption{Mapping between the suppression factors $S(k)$ across different cosmologies including massive neutrinos. The left, middle, and right columns correspond to different values of the total neutrino mass, $\Mnu=0.1,\,0.5,\,1\,\eV$, respectively. \emph{Top:} suppression factor for the cosmologies C0, C1, and C2, at 8 different values of the ratio $\sigtwelvestar/\sqrt{\As}$, specifically the ones corresponding to the C0 cosmology at redshift from $z=0.5$ (largest suppression) to $z=4$ (smallest suppression) with a step of $\Delta z=0.5$. \emph{Bottom:} fractional percent difference between the suppression factors of cosmology C1 and C2 with respect to C0.}
    \label{fig:mapping_of_suppression_factor}
\end{figure*}

Before proceeding, a few considerations are necessary. First, the value of $\sigtwelve$ used to track the evolution of the density field should be the one corresponding to the cosmology $\shapestar$, i.e., $\sigtwelvestar$, regardless of the actual neutrino mass. This ensures that the large-scale limit of $\Pstar$ matches that of the cosmology including massive neutrinos, as can be seen in Fig. \ref{fig:supp_factor}. Second, the value of $\sigtwelvestar$ alone does not fully determine the severity of neutrino damping on scales $\lambda \lesssim \lams$. This becomes evident when comparing two cosmologies for which $\Pstar$ can be mapped from one to the other through evolution mapping (and thus they share the same shape parameters, including $\Mnu$, and same $\sigtwelvestar$) and observing that the amplitude of the damping generally differs between them. This indicates that a subset of evolution parameters carries additional information beyond $\sigtwelvestar$. Considering the suppression factor $S(k)$, we note that it is independent from the scalar amplitude $\As$ since the latter has the same effect on both $P$ and $\Pstar$. This suggests that a more natural parameter to describe the evolution of the suppression factor is the ratio $\sigtwelvestar/\sqrt{\As}$, which removes the explicit dependence of $\sigtwelvestar$ on $\As$. This choice can be further justified by the fact that fixing $\sigtwelvestar$ while varying $\As$ effectively requires adjusting the redshift to match the large-scale amplitude. Consequently, a lower (higher) value of $\As$ corresponds to a lower (higher) redshift, which in turn leads to stronger (weaker) suppression, as neutrino free-streaming has had more (less) time to dampen structure formation.

To validate the preceding arguments, Fig. \ref{fig:mapping_of_suppression_factor} illustrates the suppression factor for three distinct cosmologies (C0,\,C1,\,C2), with their respective parameters detailed in Tab.~\ref{tab:cosmologies}. As with the original formulation excluding massive neutrinos, the shape parameters, this time also including the total neutrino mass, are kept constant to maintain the shape of the power spectrum. For the evolution parameters, we adopt a \Planck-like \lcdm cosmology \citep{Planck2018} as the reference model (C0), and consider two additional cases (C1 and C2) that feature significantly different values of $\As$, and the dark energy equation of state $w$ and energy density $\omde$. By evaluating the rescaling of the suppression factor across three different values of $\Mnu$, this approach effectively explores a total of nine distinct cosmologies. 


\begin{table}
    \centering
    \caption{Cosmological parameters assumed for the reference cosmology C0 and two other configurations (C1,\,C2). In order, different rows show the value of the baryon $\omb$ and total matter $\omm$ density parameter, the scalar index $\ns$, the dark energy density $\omde$, the expansion rate $h$ (this is a derived parameter following Eq. \ref{eq:sum_densities}), the scalar amplitude $\As$ (normalised by the value assumed in C0), the equation of state parameter ($w$) for dark energy, and the values of the total neutrino mass ($\Mnu$). Given the requirements that the total matter density stays the same, the use of three different values for $\Mnu$ implies the effective use of nine different cosmologies.}
    \label{tab:cosmologies}
    \vspace{0.2cm}
    \renewcommand{\arraystretch}{1.3}
    \begin{tabular}{c@{\hskip 0.4cm}c@{\hskip 0.4cm}c@{\hskip 0.4cm}c}
        \hline
        \hline
        & C0 & C1 & C2 \\
        \hline
        \hline
        $\omb$ & 0.0219961 & 0.0219961 & 0.0219961 \\
        \hline
        $\omm$ & 0.1431991 & 0.1431991 & 0.1431991 \\
        \hline
        $\ns$ & 0.97 & 0.97 & 0.97 \\
        \hline
        \hline
        $\omde$ & 0.3057009 & 0.6668009& 0.1068009\\
        $h$ & 0.67 & 0.90 & 0.50 \\
        \hline
        $(\As/2.1)\,[10^{-9}]$ & 1.00 & 3.00 & 0.33 \\
        \hline
        $w$ & -1.0 & -1.1 & -0.9 \\
        \hline
        \hline
        $\Mnu\,[\eV]$ & (0.1, 0.5, 1.0) & (0.1, 0.5, 1.0) & (0.1, 0.5, 1.0) \\
        \hline
        \hline
    \end{tabular}
\end{table}


The top panels of Fig. \ref{fig:mapping_of_suppression_factor} shows the suppression factors of the cosmologies (C0,\,C1,\,C2) for the three selected values of $\Mnu$ and for eight different values of the ratio $\sigtwelvestar/\sqrt{\As}$. For all of these values, we find an exquisite match among the suppression factor of the three different cosmologies, despite the extremely different evolution parameters that they feature. To better appreciate the level of agreement, in the bottom panels we show the fractional percentage difference of C1 and C2 with respect to the reference case C0. For all the considered values of $\Mnu$, the residuals are well below the $0.01\%$ threshold, with the only exception represented by the case at $\Mnu=1\,\eV$ and for the largest value of $\sigtwelvestar/\sqrt{\As}$. Looking into the details, we can observe a trend of increasing residuals for larger value of $\sigtwelvestar/\sqrt{\As}$, as highlighted by the lighter color shades. This suggests that, while the overall agreement is strong, there may be subtle effects from variations in the remaining evolution parameters that are not fully captured by the ratio 
$\sigtwelvestar/\sqrt{\As}$. Nevertheless, the amplitude of the residual is still extremely small, and never exceeds the order of magnitude of ${\cal O}(0.01\%)$.  

In conclusion, these observations on the suppression factor $S(k)$ enable us to extend the original formulation of evolution mapping, as given by Eq. (\ref{eq:orig_evmap}), to scenarios that include massive neutrinos. The updated equation can then be expressed as
\begin{equation}
    \Plin(k\,|\,\shape,\,\evolution,\,z) = \Plin(k\,|\,\shape,\,\As,\,\sigtwelvestar(\shapestar,\,\evolution,\,z))\,.
    \label{eq:mapping_with_neutrinos}
\end{equation}
In this new formulation, $\Mnu$ is now incorporated into the shape parameters $\shape$, the scalar amplitude $\As$ must be explicitly specified and can no longer be treated as part of the remaining evolution parameters 
$\evolution$, and the time variable $\sigtwelve$ must be calculated based on the cosmology with $\shapestar$ rather than $\shape$. The requirement to treat $\As$ and $\sigtwelvestar$ independently arises because, while $S(k)$ can be mapped across different cosmologies using the ratio $\sigtwelvestar/\sqrt{\As}$, $\Pstar(k)$ requires a specific value of $\sigtwelvestar$ to correctly recover the large-scale amplitude of density fluctuations, effectively making necessary the specification of both parameters.

\subsection{Effect on the total matter power spectrum}
\label{sec:total_matter_pk}

In the previous subsection, the validation of the mapping across different cosmologies has been conducted under the implicit assumption that the power spectrum quantifies the amplitude of perturbations of \cdm and baryons. This assumption stems from the conventional approach of using a two-fluid model \citep{BlaGarKon2014, ZenBelVil2016, ZenBelDos2018} to describe the matter density field, 
$\deltacb$ and $\deltanu$. Within this framework, it is typically assumed that the clustering of biased tracers can be fully characterized by $\deltacb$ alone \citep{VilMarVie2014, CasSefShe2014, VagBriArc2018}. As a result, the relevant matter power spectrum to start from in the modelling of the galaxy power spectrum is $\Pcb\propto\langle\deltacb^{\,2}\rangle$.

\begin{figure}
    \centering
    \includegraphics[width=\columnwidth]{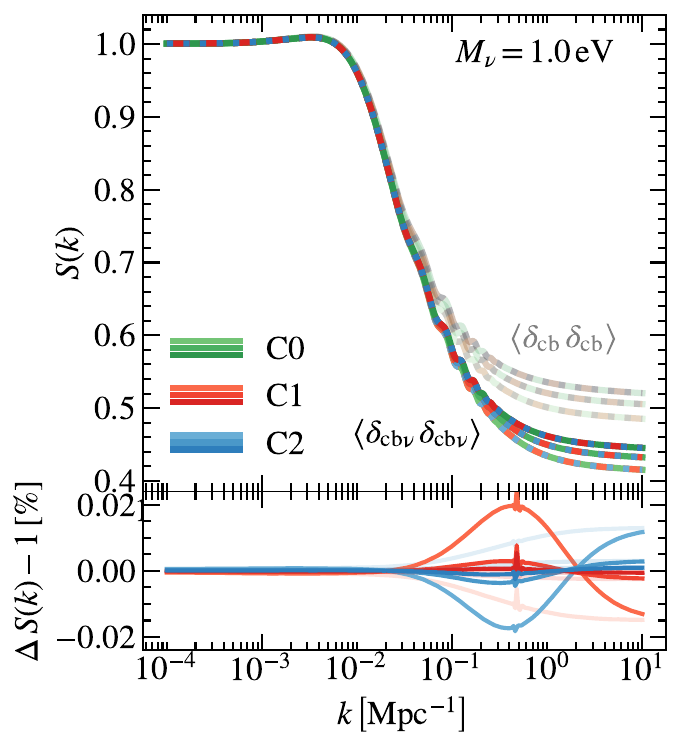}
    \caption{Same as in Fig. \ref{fig:mapping_of_suppression_factor} but displaying the comparison between the suppression factor of the c+b power spectrum (faint) and the one of the total matter component, i.e. c+b+$\nu$ (bright). In this case we only show the configuration with the largest value of the total neutrino mass, i.e. $\Mnu=1\,\eV$, and only three values for the ratio $\sigtwelvestar/\sqrt{\As}$.}
    \label{fig:diff_nonu_vs_tot}
\end{figure}

Even if the previous assumption is well justified in the limit of a small neutrino mass and for typical survey volumes from Stage-III experiments \citep{NorAviFro2022, MorTseCar2023}, its applicability to future surveys must be carefully assessed. For this reason we are interested in understanding how the evolution mapping of the suppression factor behaves when replacing $\Pcb$ with the total matter power spectrum $\Pcbnu$. In terms of the transfer function, this can be written as
\begin{equation}
    \Tcbnu(k) = \fcb\,\Tcb(k) + \fnu\,\Tnu(k)\,,
\end{equation}
where $\fcb\equiv(\omc+\omb)/\omm$ and $\fnu\equiv\omnu/\omm$ are the fractions of \cdm plus baryons, and neutrinos, respectively. In Fig. \ref{fig:diff_nonu_vs_tot} we show a comparison between the suppression factor in the two cases, still making use of the same set of cosmologies considered in Fig. \ref{fig:mapping_of_suppression_factor}. Since we only consider the case with $\Mnu=1\,\eV$, we find a significantly stronger ($\sim10\%$) suppression when considering fluctuations of the total matter density field. Nevertheless, the mapping across different set of evolution parameters still follows degeneracies for constant values of the ratio $\sigtwelvestar/\sqrt{\As}$, as highlighted in the bottom panel. We observe a larger mismatch at the transition scale where the damping factor starts to flatten out, at approximately $k\sim0.5\,\kMpc$, but still find the different curves to be consistent at better than $0.02\%$ for all the considered cases. We therefore conclude that evolution mapping can also be applied when considering the total matter power spectrum $\Pcbnu$, even if for the remainder of this article we will still reference only the \cdm plus baryons component.

\subsection{Effect on the number of massive neutrinos}

One of the key insights in neutrino cosmology is that the total mass $\Mnu$ of the different neutrino flavors is the primary parameter influencing the suppression of power below the free-streaming scale. However, subtle yet significant differences can arise in the matter power spectrum depending on the individual mass fractions $m_{\nu,i}$ of each neutrino eigenstate. This occurs because, although the total mass $\Mnu$ determines the composition and expansion rate of the Universe, variations in the mass distribution among the three neutrino flavors can cause shifts in the epoch at which they transition from a relativistic to a non-relativistic behaviour. In the previous section, we assumed a single massive eigenstate encompassing the total mass $\Mnu$, i.e. $\Nnu=1$. Here, we aim to investigate whether the mapping of the suppression factor is sensitive to this assumption or not.

\begin{figure}
    \centering
    \includegraphics[width=\columnwidth]{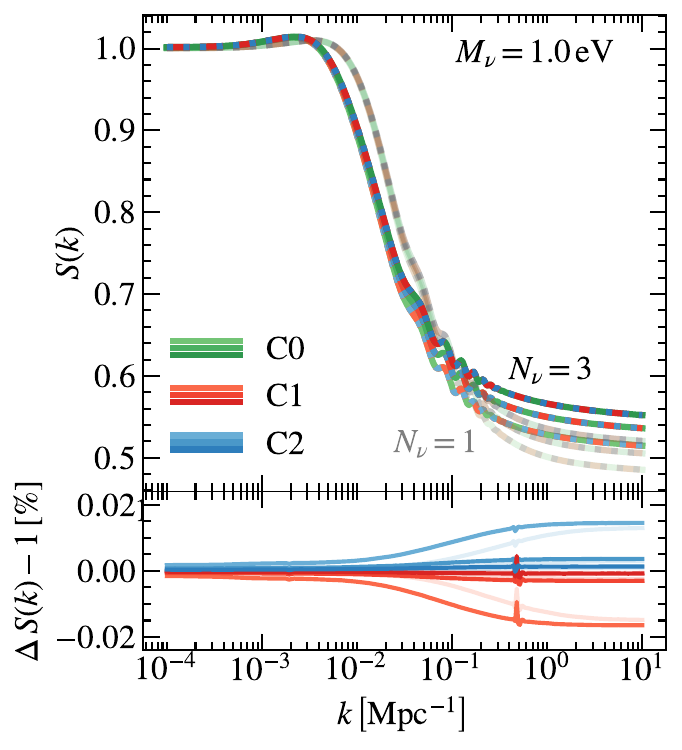}
    \caption{Same as in Fig. \ref{fig:mapping_of_suppression_factor} but displaying the comparison between the suppression factor when considering one massive neutrino flavour (faint) or three individual neutrino species with the same total neutrino mass (bright). In this case we only show the configuration with the largest value of the total neutrino mass, i.e. $\Mnu=1\,\eV$, and only three values for the ratio $\sigtwelvestar/\sqrt{\As}$.}
    \label{fig:diff_1_vs_3}
\end{figure}

In Fig. \ref{fig:diff_1_vs_3}, we compare the suppression factor as we did in Fig. \ref{fig:diff_nonu_vs_tot}, but this time we examine two cases with the same total mass $\Mnu=1\,\eV$ and different numbers of massive neutrino species: $\Nnu=1$ and $\Nnu=3$. In the second case, the mass is equally distributed among the three different eigenstates, with 
$m_{\nu,i}\simeq0.33\,\eV$ for $i=1,2,3$. In both configurations, we set the number of massless neutrino species such that the total effective number of neutrino species remains fixed at $\Neff=3.044$. This corresponds to $\Nur=2.029$ for $\Nnu=1$ and $\Nur=0$ for $\Nnu=3$. As anticipated, the latter configuration exhibits a larger free-streaming scale, since the individual neutrino masses are smaller and thus remain relativistic for a longer period. However, this also means that they begin to suppress the growth of structure later in cosmic history and for a shorter duration, resulting in a milder suppression of small-scale power compared to the configuration with a single, more massive species.

When assessing the accuracy of the evolution mapping, we observe that the three different cosmologies (C0,\,C1,\,C2) can still be effectively mapped onto each other, maintaining the high level of agreement previously seen in the $\Nnu=1$ configuration. The most notable difference is that discrepancies now begin to appear at lower $k$ values, although they eventually converge with the original differences as we approach the small-scale plateau of the suppression factor. Even in this scenario, the mapping accuracy remains better than ${\cal O}(0.02\%)$. 

We explicitly tested that these results remain consistent with more realistic neutrino mass hierarchies -- normal ($m_{\nu,1}\approx m_{\nu,2} < m_{\nu,3}$) and inverted ($m_{\nu,1} < m_{\nu,2}\approx m_{\nu,3}$) -- in addition to the degenerate case shown in Fig.~\ref{fig:diff_1_vs_3}. For these tests, we assumed a minimum total neutrino mass  of $\Mnu=0.06\,\eV$, consistent with current neutrino oscillation experiments, and considered variations of this value up to $\Mnu=0.2\,\eV$. The individual mass splits are chosen according to the assumed hierarchy: for a normal hierarchy, we adopt mass fractions of $(0.1, 0.1, 0.8)$, while for an inverted hierarchy, we use 
$(0.1,0.45,0.45)$. In all cases, we find that the residuals in the mapping do not exceed the 0.005\% limit, essentially resulting in the same matter power spectrum. These findings underscore the robustness of the suppression factor mapping across different neutrino mass distributions, confirming that, even with variations in the number of massive neutrino species and different neutrino hierarchies, the mapping achieves remarkable precision, ensuring its applicability across a wide range of cosmological models.

\section{Theory models}
\label{sec:models}


To introduce some of the theoretical quantities that are going to be relevant later in Sect. \ref{sec:emulator_design}, in this section we provide a brief overview of the two models implemented in the current version of \comet, the \eft and \vdg models \citep[see][for a more detailed discussion of their formulation and parametrisation]{EggLeeSco2024}.

In both cases, we model the overdensity of biased tracers, $\deltag$, using a standard galaxy bias expansion up to third order in the matter density $\delta$ \citep{McDRoy2009, AssBauGre2014, DesJeoSch2018, EggScoSmi2019},
\begin{equation}
    \delta_{\rm g} = \;\bone \delta + \bnablatwo\nabla^2\delta + \varepsilon_{\rm g} + \frac{\btwo}{2}\delta^2 + \gtwo\, {\cal G}_2 + \gtwoone\, {\cal G}_{21} + \ldots\,.
    \label{eq:galaxy_density_field}
\end{equation}
This expansion incorporates both local ($\delta, \delta^2$) and non-local (${\cal G}_2, {\cal G}_{21}$) operators \citep{ChaScoShe2012,EggScoSmi2019}, parameterized by four bias parameters: the linear bias 
$\bone$, the quadratic bias $\btwo$, and the leading- and next-to-leading-order tidal biases, $\gtwo$ and $\gtwoone$. Additionally, the model includes a higher-derivative contribution $\nabla^2\delta$, modelling spatial non-localities in the formation of galaxies, and a stochastic correction $\varepsilon_{\rm g}$ to account for the effects of short-wavelength fluctuations on galaxy clustering at larger scales. As already discussed in Sect. \ref{sec:total_matter_pk}, in cosmological models that contain massive neutrinos, we define all operators in the bias relation with respect to fluctuations in the \emph{cold} matter density, i.e., the combination of the \cdm and baryon components. In contrast to a definition based on the total matter density, which includes the neutrino fluctuations, this prescription has been observed to yield a halo mass function that is close to universal and a linear bias parameter that is only weakly scale-dependent in simulations \cite{VilMarVie2014,CasSefShe2014,CasCarBel2015}. The scale-dependent growth in massive neutrino cosmologies generally introduces a scale-dependence in the galaxy bias at the few percent level \cite{HuiPar0802,ParHuiShe1103}. Since this scale-dependent signal would be difficult to disentangle from the nonlinear and higher-derivative contributions to the bias relation, we do not consider it further in our modelling.


In redshift space, under the assumptions of mass conservation in the mapping from real- to redshift-space coordinates and the distant-observer approximation, the anisotropic galaxy power spectrum for a wave mode $k$ and an orientation to the line of sight $\theta$ (with $\mu\equiv\cos(\theta)$) can be written as \citep{ScoCouFri1999}
\begin{equation}
    \begin{split}
        \Pggs(k,\mu) = &\int \frac{{\rm d} {\rv}}{(2\pi)^3} \,e^{i\kv\cdot\,\rv} \, \bigg\langle e^{\lambda \,\Delta\upara} \,\times \\
        & \hspace{3pt}\left[\deltag(\xv) +f\,\nablapara\upara(\xv)\right] \left[\deltag(\xv')+f\,\nablapara\upara(\xv')\right] \bigg\rangle \,,
    \end{split}
    \label{eq:power_zspace}
\end{equation}
where $\rv\equiv\xv\,-\,\xv'$, $\Delta\upara\equiv\upara(\xv)\,-\,\upara(\xv')$, and $\lambda=-ifk\mu$. The quantity $\uv$ represents the normalized velocity field, defined as 
${\uv\equiv-\vvv\,/\,[f\,a\,H(a)}]$, such that 
$\nabla\cdot\uv=\delta$ in linear perturbation theory. In addition, $a$ and $H(a)$ denote the scale factor and the Hubble parameter, respectively, the subscript ``$\parallel$'' refers to components along the line of sight, and the growth rate 
$f$ is given by the logarithmic derivative of the linear growth factor $D(a)$,
\begin{equation}
    f(a) \equiv \frac{{\rm d}\log D(a)}{{\rm d}\log a}\,.
\end{equation}
In cosmologies including massive neutrino, we assume that galaxy velocities, like the galaxy density, are sourced by the \cdm and baryon components, consistent with findings from simulations \cite{VilBanDal1807,VerBelMor2503}. The growth rate is accordingly computed for the \cdm+baryon fluid, but we additionally neglect any scale-dependence induced by neutrinos and account for their suppression effect only through the linear power spectrum.\footnote{We make use of the total matter density (\cdm+b+$\nu$) to predict the large-scale limit of the growth rate, as this quantity correctly reproduces the large-scale limit of the ratios $\Pdt/\Pdd$ and $\Ptt/\Pdt$.} This approximation has been found to be accurate for sufficiently small neutrino masses, $M_{\nu} \lesssim 0.1\,\mathrm{eV}$ \cite{ChuIva1911,AviBanNiz2111}, but an exact treatment may be required when exploring larger $M_{\nu}$ values as allowed by analyses that consider galaxy clustering data without the addition of \cmb information \cite{NorAviGil2501}.

The expression for the redshift-space power spectrum in Eq.~(\ref{eq:power_zspace}) reveals two distinct regimes of distortions. The terms in square brackets describe large-scale power enhancement due to coherent infall of galaxies into overdense regions \citep{Kaiser1987, Hamilton1992}, while the velocity difference  generating function (VDG), defined as
\begin{equation}
    W(\lambda,\rv)\equiv \left\langle e^{\lambda\Delta u_\parallel}\right\rangle\,,
\end{equation}
suppresses small-scale power via the \fog effect \citep{Jackson1972}, caused by  velocity dispersion inside virialised structures. In practice, these regimes are tightly intertwined, with no sharp transition scale. 


In  \citep{Sco2004} it was shown that the expectation value in Eq.~(\ref{eq:power_zspace}) can be factorized, with the VDG being a prefactor that controls the small-scale damping of the power spectrum multipoles, and a second factor that depends on cross-correlations with the (large-scale) density perturbations, which can be expanded in perturbation theory up to e.g. one-loop~\citep{Sco2004, TarNisSai2010,SanScoCro2017}. A more refined version of this approach is represented by the \vdg model (see~\citep{EggLeeSco2024} for recent detailed comparison against simulations), which resums quadratic non-linearities and, in the large-separation limit, $\rv\rightarrow\infty$, results in the following damping factor,
\begin{equation}
    W_\infty(\lambda) = \frac{1}{\sqrt{1-\lambda^2\avir^2}}\exp\left(\frac{\lambda^2\sigv^2}{1-\lambda^2\avir^2}\right)\,.
    \label{eq:Winfty}
\end{equation}
Here, $\avir$ is a free parameter that accounts for nonlinear contributions to velocity dispersion and controls the extended tails of the pairwise velocity distribution (i.e. its kurtosis, see e.g. \citep{BiaChiGuz2015}), and $\sigma_v$ is the bulk velocity in linear theory,
\begin{equation}
    \sigv^2 \equiv \frac{1}{6\pi^2} \int \Plin(k)\,{\rm d}k\,,
    \label{eq:sigmav}
\end{equation}
where $\Plin(k)$ is the linear matter power spectrum. Eq.~(\ref{eq:Winfty}) has been used already in analysis of galaxy redshift surveys, see e.g. \citep{SanScoCro2017, GriSanSal2017, SemSanPez2022, SemSanPez2023}.

On the other side, the second class of \rsd models describe the velocity difference generating function by expanding it in a power series, aligning with the recently-developed \eft framework \citep{CarHerSen2012, SenZal2014}. This approach introduces effective speed-of-sound parameters to account for the small-scale \rsd damping, but are also meant to describe the impact of shell crossing and the breakdown of perturbation theory for ultraviolet modes. At leading order, the non-virial power spectrum includes terms of the form $k^2\Plin(k)$ and even powers of $\mu$,
\begin{equation}
    \Pctrlo(k,\mu) = -2 k^2 \Plin(k) \sum_{i=0,2,4}c_{i}\,{\cal L}_{i}(\mu)\,, \label{eq:Pctr_lo}
\end{equation}
where ${\cal L}_\ell$ is the $\ell$-th order Legendre polynomial, and $\czero,\ctwo,\cfour$ parametrise the leading \rsd contributions to the multipoles $\ell=0,2,4$. To account for stronger non-linearities in the real-to-redshift mapping, next-to-leading-order corrections are often included,
\begin{equation}
    \Pctrnlo(k,\mu) = \cnlo k^4 \mu^4 f^4 \left(\bone+f\mu^2\right)^2 \Plin(k)\,, \label{eq:Pctr_nlo}
\end{equation}
where $\cnlo$ is an additional nuisance parameter to be marginalized over in likelihood analyses.

The final expressions for the models discussed above, incorporating biased tracers, can be summarized as follows. In the \eft framework, the redshift-space galaxy power spectrum is written as
\begin{equation}
    \begin{split}
        &\PggEFT(k,\mu) = \\
        &\hspace{15pt}\Pggtree(k,\mu) + \Pggoneloop(k,\mu) + \Pggctr(k,\mu) + \Pggnoise(k,\mu)\,,
    \end{split}
\end{equation}
where tree-level and one-loop terms arise from the coupling of the density and velocity fields, $\Pggctr$ includes the counterterms from Eqs. (\ref{eq:Pctr_lo}) and (\ref{eq:Pctr_nlo}), and $\Pggnoise$ accounts for stochastic contributions. The latter, modeled perturbatively and truncated at next-to-leading order, can be written as
\begin{equation}
    \Pggnoise(k,\mu) = \frac{1}{\bar{n}}\left[\npzero + k^2\left(\nptwozero + \nptwotwo\,{\cal L}_2(\mu)\right)\right]\,, \label{eq:shot_noise_contribution}
\end{equation}
where $\bar{n}$ is the mean number density of the sample, and $\npzero,\nptwozero, \nptwotwo$ are free parameters describing deviations from Poisson noise.

Using the same formalism, the \vdg model can be written as
\begin{equation}
    \begin{split}
        \PggVDG&(k,\mu) = \\
        &\Winfty(\lambda) \, \Big[\Pggtree(k,\mu) + \Pggoneloop(k,\mu) -\Delta P(k,\mu) \\
        & + \Pctrlo(k,\mu)\Big] + \Pggnoise(k,\mu)\,.
    \end{split}
\end{equation}
A few important clarifications are needed here. First, in the \vdg model, certain additional contributions that appear at one-loop in the \eft model are not required. This is because they are implicitly accounted for by treating the velocity difference generating function analytically, without expanding it in power series as in the \eft model. Consequently, the one-loop term in the \vdg model can be recast in term of the \eft one by subtracting a correction of the form \citep{EggLeeSco2024}
\begin{equation}
    \begin{split}
        \Delta P(k,\mu) = &\int \frac{{\rm d}\rv}{(2\pi)^3} e^{i\kv\cdot\rv} \, \frac{\lambda^2}{2}\Big\langle\Delta \upara^2\Big\rangle \\
        &\hspace{2pt}\times \Big\langle\left[\delta(\xv) +f\nablapara\upara(\xv)\right] \left[\delta(\xv')+f\nablapara\upara(\xv')\right]\Big\rangle\,.
    \end{split}
\end{equation}
Secondly, the \rsd damping described by the analytical function $\Winfty$ does not apply to the stochastic term $\Pggnoise$ \citep{EggLeeSco2024}, such that stochastic corrections are equivalent across the two \rsd models. Finally, we notice that despite the small-scale \rsd signal being modelled through $\Winfty$, the \vdg model still features the leading-order counterterms present in the \eft formalism. This choice is motivated to correctly capture residual effects that are equally shared with the \eft model, such as higher derivatives in the galaxy bias expansion, velocity bias, and residual breaking of perturbation theory for ultraviolet modes.

A final common feature in both models is the smearing of the \bao features caused by large-scale bulk motions \citep{CroSco2006b, EisSeoWhi2007, EisSeoSir2007, MatPie2008, CroSco2008}. A common approach to deal with this \citep{BalMirSim2015, BlaGarIva2016, IvaSib2018} involves decomposing the power spectrum into smooth ($\Pnw$) and wiggly ($\Pw$) components,
\begin{equation}
    \Plin(k) = \Pnw(k)+\Pw(k)\,.
\end{equation}
The \bao damping is then modeled by applying a damping factor to $\Pw$, yielding the \ir-resummed power spectrum at leading order,
\begin{equation}
    P_{\rm lin,IR}(k) = \Pnw(k) + e^{-k^2\Sigma^2}\Pw(k)\,, \label{eq:IR-LO}
\end{equation}
where $\Sigma^2$ represents the variance of the large-scale displacement field, which, in redshift space, receives additional anisotropic contributions. To incorporate \ir effects into the galaxy power spectrum, Eq. (\ref{eq:IR-LO}) can be used in place of $\Plin$ in all the relevant terms at both leading and next-to-leading order.

\section{Design of the emulator}
\label{sec:emulator_design}

In this section, we proceed with the description of how \comet\hspace{-1pt}\vtwo extends the design originally implemented in the first public release of \comet \citep{EggCamPez2022}. The basic structure of the code has been maintained to allow for a fast and accurate prediction of the galaxy power spectrum multipoles $\Pell(k,z)$ according to different modelling frameworks. In the next few paragraphs we describe the main modelling choices of \comet\hspace{-1pt}\vtwo, explicitly pointing out the extra features that have been added with respect to \comet. 

\subsection{Flowchart}

Similarly to the original version, \comet\hspace{-1pt}\vtwo builds upon the concept of evolution mapping, now extended to account for the influence of relic neutrinos with non-zero mass, as detailed in Sect. \ref{sec:evolution_mapping}. Since the mapping between different evolution parameters can be applied without introducing significant inaccuracies, it remains valid for a broad class of non-linear models that depend solely on the shape and amplitude of the linear power spectrum. This is typically the case for perturbative approaches, such as the \eft and \vdg models, where non-linear corrections $P_{\,\rm loop}$ can be computed via integrals of the form
\begin{equation}
    P_{\,\rm loop}(\kv) \propto \int_\qv K(\qv,\,\kv-\qv)\,\Plin(q)\,\Plin(|\kv-\qv|)\,,
    \label{eq:loop_integrals}
\end{equation}
with $\qv$ being a loop variable to be integrated over, and $K(\qv_1,\qv_2)$ representing a generic kernel describing the non-linear coupling between modes $\qv_1$ and $\qv_2$. Assuming that the kernel $K$ is independent of cosmology and that it can be accurately modeled under the \eds approximation \citep[see e.g.][for works aimed at improving over this ansatz]{Taruya2016, AmiMarPie2021, GarLaxSco2023, HarGar2023}, any mapping applied to $\Plin$ will naturally propagate to $P_{\,\rm loop}$. The only additional components required by \rsd models are the growth rate $f$ and the bulk velocity dispersion $\sigv$ (specifically for the \vdg model), both of which can still be derived in terms of $\Plin$.

For the implementation of the two \rsd models developed in the \comet package, the impact of massive neutrinos is only considered at the linear level. This is done by including damping above the neutrino suppression scale in the linear matter power spectrum, while assuming that the functional form of the mode-coupling kernels from perturbation theory remains consistent with the massless neutrino scenario. Although there are studies that investigate the theoretical systematics introduced by this approximation \citep[see e.g.][]{SaiTakTar2009, BlaGarKon2014, CasCarBel2015, ParCarBel2022}, addressing these effects is not the focus of this paper and will be considered in a future version of the code.

The approach of \comet\hspace{-1pt}\vtwo remains consistent with the methodology employed in the previous version. To avoid training the emulator directly on the bias and \rsd parameters (introducing 11 additional dimensions in the input parameter space of the emulator), we take the approach of emulating the individual terms of the one-loop expansion, and reconstruct the final power spectrum a posteriori. Specifically, all contributions to the power spectrum multipoles, $P_{{\cal B},\ell}$, arising from the non-linear expansion, as described in Eqs. (\ref{eq:Pctr_lo}-\ref{eq:Pctr_nlo}) and (\ref{eq:galaxy_density_field}), are organized according to their distinct combinations of bias and counterterm coefficients $\cal{B}$. This categorization allows for the emulation of a reduced set of independent terms, which can subsequently be analytically recombined to construct a model for $\Pell(k,z)$. To reduce the dynamical range of the training data, therefore improving the accuracy of the emulation, the raw $P_{{\cal B},\ell}$ are not directly fed into the training routines. Instead, they are first rescaled by the corresponding linear power spectrum, so that the emulator outputs the normalized quantity $\beta_{{\cal B},\ell} \equiv P_{{\cal B},\ell} / \Plin$.  

Altogether, there are 19 individual terms per multipole, which are outlined in Tab. \ref{tab:emulated_terms} as products of the model parameters described in Sect. \ref{sec:models}. The list does not explicitly mention any of the stochastic parameters from Eq. (\ref{eq:shot_noise_contribution}) because the terms associated to the stochastic field are not emulated; instead, they are generated analytically and summed to the anisotropic galaxy power spectrum a posteriori. Similarly, other analytical effects can be applied afterwards, such as the small-scale damping $\Winfty$, resulting in a reconstructed 2D spectrum given by
\begin{equation}
    \begin{split}
        \Pgg(k,\mu) = & \;\Winfty(k,\mu,\sigv) \, \left[\sum_{\ell} \legell(\mu) \sum_{{\cal B}}{\cal B}\,P_{{\cal B},\ell}(k)\right] \\ 
        & + \Pggnoise(k,\mu)\,,
    \end{split}  
    \label{eq:reconstructed_pkmu}
\end{equation}
where $\Winfty$ corresponds to Eq. (\ref{eq:Winfty}) for the \vdg model or is equal to 1 for the \eft model.

\begin{table}
    \centering
    \caption{Complete list of coefficients ${\cal B}$ for the terms emulated by \comet. This is split into those terms that enter at leading (tree) or one-loop order, and further that are necessary for the real-or redshift-space galaxy power spectrum (in the latter all the terms are required). The $\bone^2$ contribution is present both at leading and next-to-leading order because we emulate the two contributions separately, to allow us to retrieve the leading-order \ir-resummed power spectrum. The $1$ in the redshift-space one-loop section represents the term sourced by the coupling of the velocity field with itself (for which there is no free parameters in the models reproduced by \comet).} The terms marked in magenta are necessary only for the \eft model.
    \label{tab:emulated_terms}
    \vspace{0.2cm}
    \renewcommand{\arraystretch}{1.5}
    \begin{tabular}{c@{\hskip 0.6cm}c@{\hskip 0.6cm}c}
        \hline
        \hline
        & {\bf Real space} & {\bf Redshift space} \\
        \hline
        \hline
        \multirow{2}{*}{{\bf Tree level}} &\multirow{2}{*}{$\bone^2$,\; $\czero$} & $\ctwo$,\; $\cfour$, \\ & & \textcolor{magenta}{$\bone^2\cnlo$},\; \textcolor{magenta}{$\bone\cnlo$},\; \textcolor{magenta}{$\cnlo$} \\
        \hline
        \multirow{2}{*}{{\bf One loop}} & $\bone^2$,\; $\bone\btwo$,\; $\bone\gtwo$,\; $\bone\gtwoone$, & \multirow{2}{*}{$\bone$,\; $1$,\; $\btwo$,\; $\gtwo$,\; $\gtwoone$} \\
        & $\btwo^2$,\; $\btwo\gtwo$,\; $\gtwo^2$ &  \\
        \hline
        \hline
    \end{tabular}
\end{table}

The transversal and longitudinal components of the wave mode $\kv$ are then distorted as
\begin{align}
    \kperp'&=\qperp\kperp\,,\\
    \kpara'&=\qperp\kpara\,,
\end{align}
to account for the impact of the fiducial cosmology assumed to convert redshifts into comoving distances. Here $\qperp$ and $\qpara$ are the \apdist parameters \citep{AlcPac1979}, defined as \citep{BalPeaHea1996}
\begin{align}
    \qperp(z) & = \frac{\DM(z)}{\DM'(z)},\\
    \qpara(z) & = \frac{\Dh(z)}{\Dh'(z)},
\end{align}
where $\DM$ and $\Dh$ are the comoving angular diameter and Hubble distances, respectively, and the prime indicates quantities in the fiducial cosmology.

\begin{figure*}
    \centering
    \includegraphics[width=2\columnwidth]{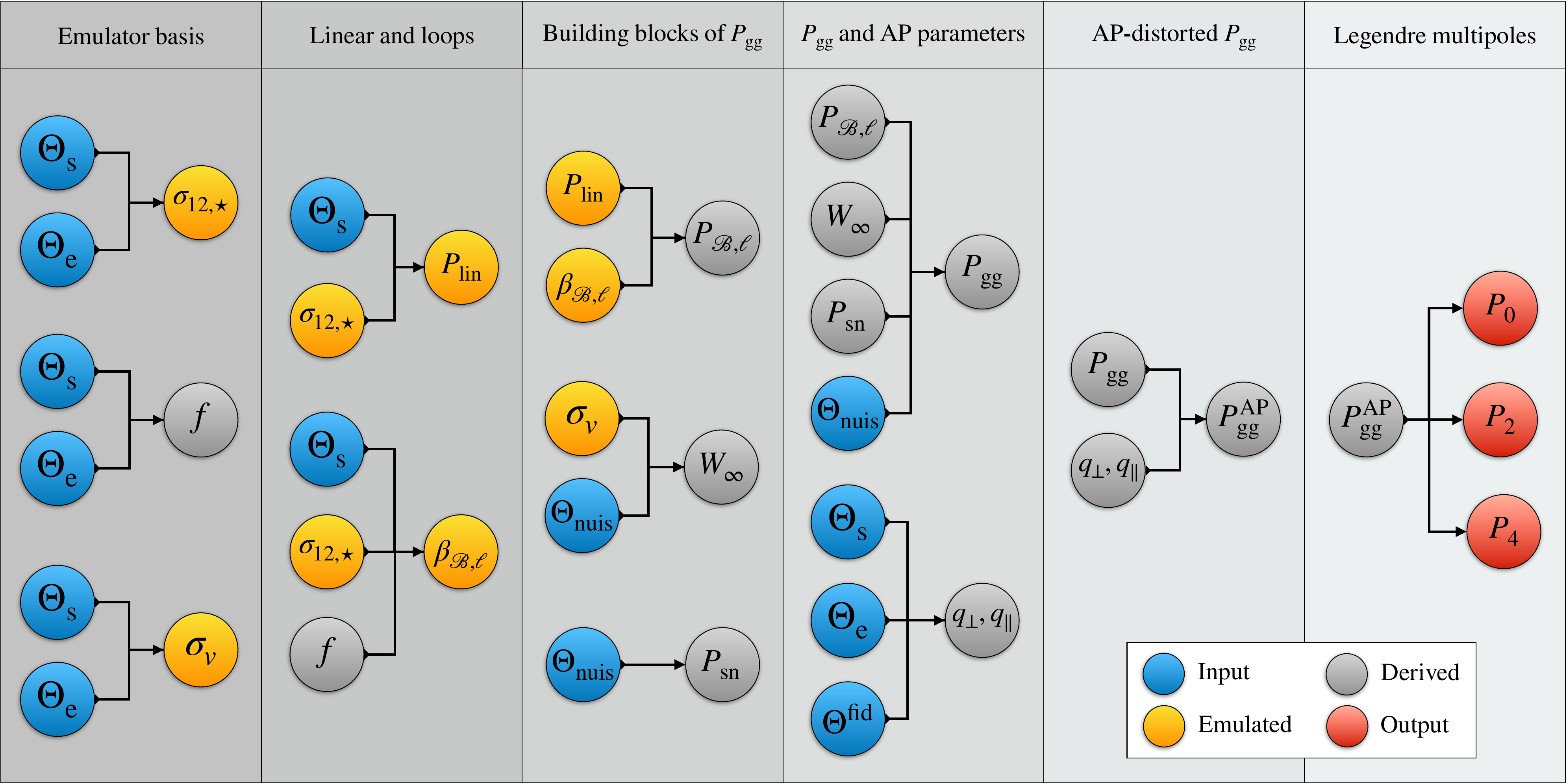}
    \caption{Flowchart illustrating the steps employed by \comet to derive the power spectrum multipoles $\Pell(k,z)$ from the initial input parameters. These inputs include shape $\shape$, evolution $\evolution$, and nuisance parameters $\nuis$, as well as the fiducial cosmology $\fiducial$ used for computing \apdist corrections. The input and output elements are marked with blue and red boxes, respectively, while all the quantities that involve the evaluation of an emulator are denoted with an orange box. Grey boxes represent quantities that can be derived analytically, without the use of any emulator. The main difference between the emulators including massive neutrinos or not is that $\As$ is either a shape or evolution parameter. In addition, the evaluation of the linear power spectrum changes in the two configurations: in one case, the output of the $\Plin$ emulator is rescaled a posteriori by the ratio of $\sigtwelve$ values as in Eqs. (\ref{eq:s12_rescaling}-\ref{eq:Plin_rescaling}), while, in the other, $\sigtwelvestar$ is requested as input for the $\Plin$ emulator itself.}
    \label{fig:flowchart}
\end{figure*}

The final step involves projecting the reconstructed anisotropic power spectrum $P(k,\mu)$ onto the Legendre polynomials $\legell$ to derive the \apdist-distorted power spectrum multipoles,
\begin{equation}
    \Pell'(k')=\frac{2\ell+1}{2\,\qperp^2\qpara}\int_{-1}^{+1} {\rm d}\mu'\, \legell(\mu')\, \Pgg\left(k(k',\mu'),\,\mu(k',\mu')\right)\,.
\end{equation}
Given that the non-linear models considered here receive contributions up to $\mu^8$ from the coupling of the density and velocity fields, the reconstruction from Eq. (\ref{eq:reconstructed_pkmu}) remains incomplete if the summation over $\ell$ only includes orders $\ell=0,2,4$. This is further reinforced by the inclusion of \ir-resummation in the model, as described in Sect. \ref{sec:ir-resummation}, which leads to the appearance of non-zero even multipoles at all orders. To account for this effect, we include also the $\ell=6$ multipole in the resummation. Since all the terms $\ell\geq6$ display a small amplitude, we opt not to create a full-fledged emulator for it, as we have done for the $\ell=0,2,4$ multipoles. Instead, we calculate a single realization of the corresponding tables using a \Planck cosmology based on the TT,TE,EE+lowE+lensing constraints \citep{Planck2018}. We then rescale each element $P_{{\cal B},6}^{\,\Planck}$ according to the values of the amplitude $\sigtwelve$ and the growth rate $f$, as computed from the evolution parameters of the cosmology being considered. The rescaling is then realised according to the following formula,
\begin{equation}
    P_{{\cal B},6} = \left(\frac{\sigtwelve(\shape,\,\evolution,\,z)}{\sigtwelve(\shapeplanck,\,\evolutionplanck,\,z=1)}\right)^{2L} P_{{\cal B},6}^{\,\Planck}\,,
\end{equation}
where $L=1$ for leading-order terms and $L=2$ for next-to-leading-order terms, as indicated in Tab. \ref{tab:emulated_terms}. As demonstrated in \cite{EggCamPez2022}, this approach effectively reduces systematic errors observed in the validation stage, particularly when applying \apdist corrections to the \vdg model. Therefore, we continue to employ this method to ensure the accuracy and robustness of our model.

Fig. \ref{fig:flowchart} provides a schematic view of the workflow used by \comet to evaluate the galaxy power spectrum multipoles. The process begins with the input parameters (marked by blue circles) and concludes with the computation of the \apdist-corrected Legendre multipoles (red circles). The emulated quantities (shown in orange) are computed as described in Sect. \ref{sec:emulators}. All other quantities (grey circles) are computed analytically, as they are quick calculations and have negligible impact on the total evaluation time of the process.

\subsection{Training and parameter priors}

\subsubsection{Gaussian processes}
The latest version of \comet has been trained using a \gp algorithm implemented via the public software \gpy.\footnote{\url{https://gpy.readthedocs.io/en/deploy/}} Gaussian processes \citep[see e.g.][for a review on the subject]{RasWil2005} are an unsupervised learning method designed to find an optimal interpolation scheme across a set of scattered $n$-dimensional points $\{\xv_i,\yv_i\}_{i=1,N}$. The accuracy of this process largely depends on the covariance function $K(\xv_i,\xv_j)$ between different input positions $\xv_i$ and $\xv_j$, which determines the relative weights assigned to the evaluation of the dependent variable $\yv$ at a new position $\xv$. Consistently with the training approach of \comet, we employ a covariance function that combines a \rbf kernel,
\begin{equation}
    K_{\rm RBF}(\xv,\xv')=\sigma_{\rm RBF}^2\exp\left[-\sum_{i=1}^d\frac{(\,\xv_i-\xv'_i)^{\,2}}{2\,\lv_{{\rm RBF},i}^{\,2}}\right]\,,
\end{equation}
with a Matérn kernel of degree $\nu=3/2$\,,
\begin{equation}
    \begin{split}
     K_{\rm Mat}(\xv,\xv')=\;&\sigma_{\rm Mat}^2\left(1+\sqrt{3}\sum_{i=1}^d\frac{(\,\xv_i-\xv'_i)^{\,2}}{\lv_{{\rm Mat},i}^{\,2}}\right)\\
     &\times\exp\left(-\sqrt{3}\sum_{i=1}^d\frac{(\,\xv_i-\xv'_i)^{\,2}}{\lv_{{\rm Mat},i}^{\,2}}\right)\,,
    \end{split}
\end{equation}
where $d$ is the dimension of the input parameter space, while $\lv$ and $\sigma$ are hyperparameters characterizing the correlation length (of different input parameters) and amplitude, respectively, at different positions $\xv$ and $\xv'$. Due to incompatibilities with the latest available versions of \numpy ($\geq{\tt 1.24}$),\footnote{\url{https://numpy.org/}} the optimal hyperparameters obtained from the training process have been reapplied to train new models using the \scikit software,\footnote{\url{https://scikit-learn.org/}} which maintains full compatibility with \numpy, allowing for seamless integration. Due to the consistency in the internal structures of both packages, this procedure was implemented efficiently, yielding machine-level precision in the comparative outputs of the two emulators.

Future plans for the \comet software include transitioning the emulators from a \gp-based architecture to a neural network. This change could reduce the evaluation time of the emulator without compromising its accuracy, leveraging the ability of neural networks to maintain performance regardless of the training set size. 

\subsubsection{List of emulators}
\label{sec:emulators}
To preserve the original accuracy level outlined in \cite{EggCamPez2022}, the initial \comet emulators are retained as distinct entities. These emulators can be selected by users who require predictions that exclude the effects of massive neutrinos. However, considering new extra features (see Sect. \ref{sec:extra_features}), such as an extended support in $k$, the extended input parameter space, and the introduction of a new \ir-resummation scheme (see Sect. \ref{sec:ir-resummation}) needed for the massive-neutrino emulators, for consistency we also decided to retrain the original configurations.

In \comet, we had two distinct groups of emulated quantities:
\begin{description}
    \item[(-) {\bf Shape}] Emulating $\Plin$, $\sigtwelve$, and $\sigv$ as functions of $\shape$ (with fixed evolution parameters $\evolutionfixed$ and $z=1$).
    \item[(-) {\bf Full}] Emulating $\beta_{{\cal B},\ell}$ as a function of $\shape$, $\sigtwelve$, and $f$.
\end{description}
Predicting $\Plin$ solely as a function of $\shape$ is feasible because, once the shape of the power spectrum is determined, the overall amplitude can be obtained using the predicted $\sigtwelve$ and then rescaling it by the ratio of the growth factors $D(z)$ and the scalar amplitudes $\As$ between the requested cosmology and the one with $\evolutionfixed$ and $z=1$ as follows:
\begin{align}
    &\sigtwelve(\shape,\evolution,z) = \nonumber\\
    &\hspace{0.9cm}\frac{D(\shape,\evolution,z)}{D(\shape,\evolutionfixed,z=1)}\left(\frac{\As}{\As^{\rm fixed}}\right)^{1/2}\sigtwelve(\shape,\evolutionfixed,z=1)\,,\label{eq:s12_rescaling}\\
    &\Plin(k|\shape,\evolution,z) = \nonumber\\
    &\hspace{1cm}\left[\frac{\sigtwelve(\shape,\evolution,z)}{\sigtwelve(\shape,\evolutionfixed,z=1)}\right]^2\Plin(k|\shape,\evolutionfixed,z=1)\,. \label{eq:Plin_rescaling}
\end{align}
A similar argument can be used to obtain predictions for the bulk velocity dispersion $\sigv$.

With the inclusion of massive neutrinos, the previous approach becomes inadequate, as $\As$ can no longer be treated as an evolution parameter. In this scenario, we can distinguish three different sets of emulators:
\begin{description}
    \item[(-) {\bf Shape}] Emulating $\sigtwelvestar$ as a function of $\shape$ (with fixed evolution parameters $\evolutionfixed$ and $z=1$).
    \item[(-) {\bf Linear}] Emulating $\Plin$ and $\sigv$ as a function of $\shape$, $\As$, and $\sigtwelvestar$.
    \item[(-) {\bf Full}] Emulating $\beta_{{\cal B},\ell}$ as a function of $\shape$, $\As$, $\sigtwelvestar$, and $f$.
\end{description}
The new mapping described in Sect. \ref{sec:evolution_mapping} relies on the $\sigtwelvestar$ computed from the equivalent cosmology with $\shapestar$. Thus, it remains feasible to predict $\sigtwelvestar$ using only the shape parameters and then rescale its value as described in Eq. (\ref{eq:s12_rescaling}). On the contrary, both $\Plin$ and $\sigv$ (which directly depends on $\Plin$ via Eq. \ref{eq:sigmav}) must now be expressed in terms of the shape parameters $\shape$, which also include 
$\Mnu$, the scalar amplitude $\As$, and the large-scale amplitude $\sigtwelvestar$. This approach allows us to accurately predict both the neutrino-free power spectrum $\Pstar$ (through $\sigtwelvestar$) and the suppression factor $S$ (through the ratio $\sigtwelvestar/\sqrt{\As}$). As a result, the total linear power spectrum can be directly written as in Eq. (\ref{eq:mapping_with_neutrinos}).

\subsubsection{Parameter ranges}
\label{sec:parameter-ranges}

\begin{table}
    \centering
    \caption{Parameters and ranges covered by the emulators in \comet. Without massive neutrinos, they include the baryon ($\omb$) and \cdm ($\omc$) density parameters, the scalar index ($\ns$), the amplitude $\sigtwelve$, and the growth rate ($f$). With massive neutrinos, it is required to specify also the total neutrino mass ($\Mnu$) and the scalar amplitude ($\As$).}
    \vspace{0.2cm}
    \label{tab:priors}
    \renewcommand{\arraystretch}{1.3}
    \begin{tabular}{c@{\hskip 0.6cm}c@{\hskip 0.6cm}c@{\hskip 0.6cm}c@{\hskip 0.6cm}c@{\hskip 0.6cm}c}
        \hline
        \hline
        \multirow{2}{*}{Parameter} & \multicolumn{2}{c}{\comet} & \multicolumn{2}{c}{\comet\hspace{-1pt}\vtwo}\\
        & min & max & min & max \\
        \hline
        \hline
    $\omb$ & 0.02050 & 0.02415 & 0.01930 & 0.02535 \\
        \hline
        $\omc$ & 0.085 & 0.155 & 0.08 & 0.16 \\
        \hline
        $\ns$ & 0.92 & 1.01 & 0.90 & 1.03\\
        \hline
        $\sigma_{12,\star}$ & 0.2 & 1.0 & 0.2 & 1.0\\
        \hline
        $f$ & 0.5 & 1.05 & 0.5 & 1.05\\
        \hline
        $\Mnu\,[\eV]$ & - & - & 0.0 & 1.0\\
        \hline
        $\As\,[10^{-9}]$ & - & - & 1.0 & 3.5\\
        \hline
        \hline
    \end{tabular}
\end{table}

As detailed in the previous sections, the parameter space spanned by the emulators differs between the original version of \comet and the new one, which also incorporates massive neutrinos. The complete set of parameters and their ranges is listed in Table~\ref{tab:priors}, explicitly highlighting the differences between the two configurations.

In the previous version of \comet, the emulator parameter space included the physical baryon density ($\omb$), the cold dark matter density ($\omc$), the scalar spectral index ($\ns$), the amplitude $\sigma_{12,\star}$, and the growth rate $f$ (the latter is not required in real space). In the updated version, the parameter ranges have been expanded: for $\omb$ and $\ns$, by almost a factor of 2 and 1.5, respectively, and slightly less so for $\omc$. These ranges now correspond to 33, 20, and 15 times the $1\sigma$ constraints reported in \cite{Planck2018} for $\omc$, $\omb$, and $\ns$, respectively.

With the inclusion of massive neutrinos, the emulator parameter space additionally includes the total neutrino mass ($\Mnu$) and the scalar amplitude ($\As$).\footnote{For this new release of \comet, we train the emulator assuming one massive and two massless neutrino species, corresponding to a total effective number of neutrino species fixed at $\Neff = 3.044$. Since changes to either the number of massive species or the value of $\Neff$ alter the shape of the matter power spectrum, any deviation from this configuration would require a dedicated training set and emulator. Such extensions will be considered in future releases of \comet.} We have adopted sufficiently broad ranges for these two parameters, particularly for $\Mnu$, which spans the interval $[0,1]\,\eV$. As highlighted by recent analyses of the combined BOSS galaxy and eBOSS quasar samples \citep{SemSanPez2023}, these limits are consistent with the $2\sigma$ confidence intervals obtained when considering extensions to \lcdm that include an evolving dark energy equation of state and non-zero curvature.

This extended and broadened parameter space necessitates a larger training dataset to achieve a similar level of accuracy as the previous version of \comet. However, instead of sampling the full parameter space as done before, we now exploit correlations between parameter combinations, such as between $f$ and $\sigma_{12,\star}$, and between $\omc$ and $\As$.  This approach allows us to exclude parts of the parameter space that are unlikely to be sampled in any realistic scenario, thereby minimizing the required increase of the training dataset size.

In practice, we proceed as follows: we begin by generating \lhc samples in the range from 0 to 1 for each dimension.  We then establish upper and lower boundaries for the growth rate as a function of $\sigma_{12,\star}$, i.e., $f_{\rm max}(\sigma_{12,\star})$ and $f_{\rm min}(\sigma_{12,\star})$, and perform inverse transform sampling to ensure a uniform sample density in the $f$-$\sigma_{12,\star}$ parameter plane. This involves computing the cumulative distribution function
  \begin{equation}
    \label{eq:CDFs12}
    \mathrm{CDF}(\sigma_{12,\star}) = {\cal N}\,\int_{0.2}^{\sigma_{12,\star}} \mathrm{d}x \, \left[f_{\rm max}(x) - f_{\rm min}(x)\right]\,,
  \end{equation}
  with normalization factor ${\cal N}$, and applying its inverse to the \lhc samples. We obtain the $\omc$ samples analogously, using upper and lower limits of $\As$ as a function of $\omc$.  Meanwhile, all other \lhc samples are scaled according to the boundaries specified in Table~\ref{tab:priors}.  The upper and lower boundaries for the growth rate are motivated by allowing a $10\,\%$ variation around $f_{\rm Planck}(z)$---calculated using the cosmological parameters of \cite{Planck2018}---and a $20\,\%$ variation around $\sigma_{12, \rm Planck}(z)$ within the redshift range of $z=0$ to $3$.  Similarly, we establish the boundaries on $\As$ as a function of $\omc$ by translating a $20\,\%$ variation in $\sigma_{12}$ at different $\omc$ values, within the range specified in Table~\ref{tab:priors}, into $\As$. For reference, the precise relationships for all these boundaries are given in Appendix~\ref{sec:app.boundaries}.


\begin{figure}
    \centering
    \includegraphics[width=\columnwidth]{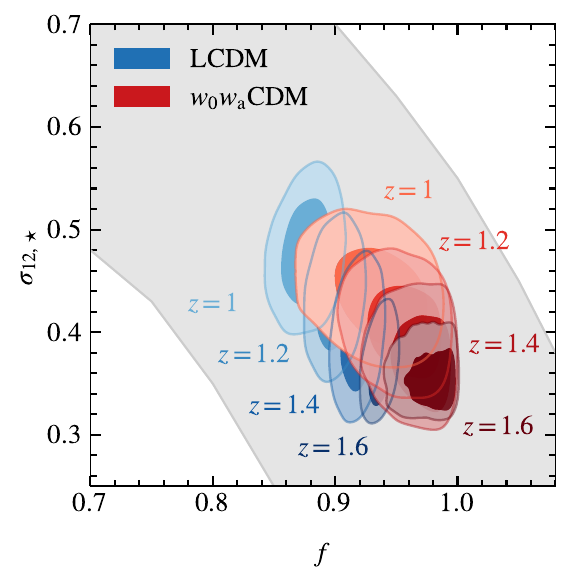}
    \caption{Distribution in the $f$-$\sigtwelvestar$ parameter space of the training set for the emulators that incorporate massive neutrinos (grey shaded area). For reference, we also show the $68\%$ and $95\%$ confidence intervals obtained from a set of points sampled during the posterior distribution exploration of four synthetic data vectors, corresponding to redshifts $z=(1.0, 1.2, 1.4, 1.6)$, as explained in the main text. The blue and red colors denote the posterior distributions for the \lcdm and \wowacdm parameter spaces, respectively. For clarity, we only show a fraction of the total $f$-$\sigtwelvestar$ plane, which also extends beyond the top left corner of the plot, as specified in Table \ref{tab:priors}.}
    \label{fig:geometry_training_set}
\end{figure}

In Fig. \ref{fig:geometry_training_set}, we present a section of the $f$-$\sigtwelvestar$ parameter space, indicated by the grey-shaded area. Additionally, we display posterior distributions, obtained when fitting synthetic data vectors at four different redshifts within the range $(1, 1.6)$. For these tests, we used Gaussian covariance matrices assuming redshift shells with an angular footprint of $2500\,\sqdeg$ and a radial width of $\Delta z=0.2$ (thus roughly corresponding to at most a quarter of the expected volume explored by Stage-IV surveys such as DESI and \Euclid). Two cosmological configurations are considered: a standard \lcdm model and a \wowacdm model, each one assuming a single massive neutrino with fixed mass $\Mnu=0.06\,\eV$. In both cases, the derived values for $f$ and $\sigtwelvestar$ fully lie within the ranges covered by the training set, demonstrating that the layout of the training dataset can be confidently restricted to the region shown in Fig. \ref{fig:geometry_training_set}. 

Despite the more efficient distribution of the samples, we choose to increase the size of the training dataset to 2000 points for the ``full'' emulators without massive neutrinos and to 3000 otherwise, in order to account for the larger parameter space. Since the dimensionality of the ``shape'' and ``linear'' emulators is lower, smaller training sets are sufficient and in these cases we use 1000 and 2000 samples, respectively.

\subsection{Additional features and improvements}
\label{sec:extra_features}

In the following subsection, we describe additional new features available in the latest version of \comet, beyond the capability to explore massive neutrino cosmologies, as discussed in the previous sections.

\subsubsection{Extended support in wavemode $k$}

\comet\hspace{-1pt}\vtwo has been trained across a wider range of wave modes $k$, ranging from
$\kmin=7\cdot10^{-4}\,\kMpc$ to $\kmax=0.5\,\kMpc$ (compared to the original maximum wave mode of $0.35\,\kMpc$). This extension should not be mistaken for an attempt to model galaxy clustering and \rsd up to these scales using the perturbative methods outlined in Sect. \ref{sec:models}. Instead, this broader range is intended to accommodate modeling scenarios that require a larger wave mode coverage, without having to resort to extrapolation. These scenarios include the convolution with the survey window function, as needed when analyzing real data samples, but also the conversion of the power spectrum multipoles into multipoles of the two-point correlation function via the Hankel transform. 

\subsubsection{Batch evaluation}
\label{sec:batch_evaluation}

The new version of \comet now allows for batch evaluation, enabling multiple parameter combinations to be processed simultaneously to obtain the corresponding power spectrum multipoles in a single call. This is particularly advantageous when using \comet to evaluate the joint likelihood across multiple redshift bins. By leveraging \scikit's built-in capabilities for batch processing, computation time can be significantly reduced. For example, a single evaluation of three power spectrum multipoles over 100 $k$ values using the \eft model without massive neutrinos takes around $11\,\mathrm{ms}$ on a single $2.6\,\mathrm{GHz}$ core machine. However, when evaluating 10 samples simultaneously, the average time is only about $30\,\mathrm{ms}$, roughly offering a speedup of a factor of three compared to running the emulator ten separate times. This effect is shown later in Fig. \ref{fig:evaluation_time}, in which we compare the total evaluation time for predicting a set of power spectrum multipoles with different emulators (\eft and \vdg, both with and without massive neutrinos). 

\subsubsection{Analytical marginalization}

In many fields that involve parameter inference from data, it is common practice to address uncertainties in nuisance parameter values by marginalizing over them. This marginalization can be performed numerically, although it may become computationally expensive when dealing with a large number of nuisance parameters.

\begin{figure}
    \centering
    \includegraphics[width=\columnwidth]{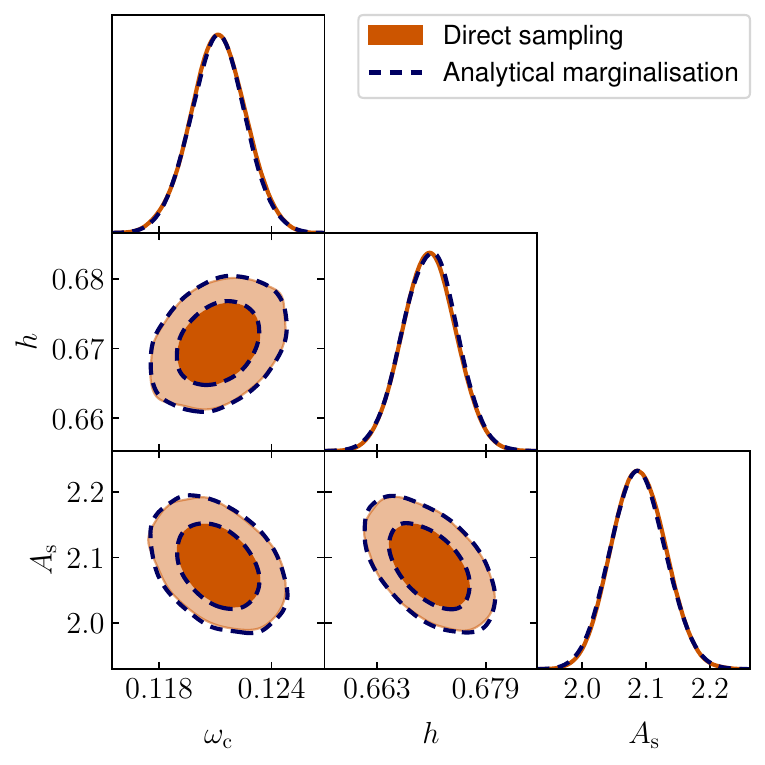}
    \caption{Comparison of the marginalised posteriors for the parameters $(\omc, h, \As)$ obtained with \comet when either explicitly sampling all linear bias, \rsd, and shot-noise parameters or marginalising over them analytically. The analysis is based on synthetic data vectors designed to replicate the clustering of H$\alpha$ galaxies, covering four redshift bins in the range $1<z<2$, and assuming a Gaussian covariance corresponding to a sky fraction of 1/3.}
    \label{fig:comp_sampled_vs_AM}
\end{figure}

A more efficient approach relies on the analytical marginalization of nuisance parameters. This method is widely applied in cosmology, with examples spanning from \cmb analyses \citep{BriCriMel2002} to \lss studies \citep{AmiSenZha2021, CarMorPou2023, RuiHadAlo2023}. Analytical marginalization can be performed under two conditions. First, the parameter to be marginalized must appear linearly in the theory model, meaning that only some of the bias and \rsd parameters in Sect. \ref{sec:models} can be treated this way. Specifically, these are the cubic non-local bias $\gtwoone$, the three leading-order and higher-order counterterms $(\czero,\ctwo,\cfour,\cnlo)$, and the three stochastic parameters $(\npzero, \nptwozero, \nptwotwo)$. Second, the prior distributions for these parameters must have a form that can be integrated analytically and for the implementation in \comet, we assume Gaussian priors.

Figure \ref{fig:comp_sampled_vs_AM} illustrates the performance of this method by comparing two chains executed with identical setups (data vectors, scale cuts, and priors), but with one chain explicitly sampling the linear nuisance parameters, and the other employing analytical marginalization. The resulting contours show excellent agreement, while the chain with analytical marginalization achieves convergence approximately ten times faster than the sampled case.\footnote{The exact speed gain of analytical marginalization is hard to quantify, since it ultimately depends on the number of parameters treated this way. Nevertheless, this approach is highly advantageous, particularly for tasks such as model selection or calibration, where multiple posterior distributions need to be computed efficiently in a short amount of time.}

\subsubsection{New \ir-resummation scheme}
\label{sec:ir-resum}

As anticipated at the end of Sect. \ref{sec:models}, large-scale bulk flows induce a smearing of the \bao wiggles, an effect that is commonly addressed by the \ir-resummation procedure. Some implementations of this procedure \cite[e.g.,][]{BalMirSim2015,BlaGarIva2016,IvaSib2018} are based on an explicit decomposition of the linear power spectrum into its smooth and wiggly components, $\Pnw$ and $\Pw$, such that $\Plin=\Pnw+\Pw$. However, there is no unique way to isolate the wiggly part, which makes the result of the \ir-resummation procedure somewhat dependent on the specific method adopted.

In the first release of \comet \citep{EggCamPez2022}, we made use of the \emph{Gaussian filter} approach, proposed by \cite{VlaSelChu2016}, which uses the featureless Eisenstein-Hu matter power spectrum \cite{EisHu1999}, $\Peh(k)$, as a template for the non-wiggly component. The non-negligible differences between the broadband shape of $\Plin$ and $\Peh$ can be accounted for by rescaling $\Peh$ with the smoothed ratio of the two power spectra \citep{VlaSelChu2016, OsaNisBer2019, PezMorZen2024, LinMorRad2024}, i.e.:
\begin{equation}
    \Pnw(k)=\Peh(k) \; {\cal F}\!\left[\frac{\Plin(k)}{\Peh(k)}\right]\,.
\end{equation}
The filter function ${\cal F}$ is chosen to be a one-dimensional Gaussian, specifically
\begin{equation}
    {\cal F}[f(k)] = \frac{\logten({\rm e})}{\sqrt{2\pi}\lambda} \int {\rm d}q \, \frac{f(q)}{q} \, \exp\left[-\frac{1}{2\lambda^2}\logten^2\left(\frac{k}{q}\right)\right] \,,
\end{equation}
where $\lambda = 0.25\,\Mpch$ is a parameter that determines the width of the filter. While this algorithm is well suited for a \lcdm cosmology, the inclusion of massive neutrinos leads to further differences in the broadband shape that can no longer be corrected for in this approach, particularly for large values of the neutrino mass, $\Mnu\sim1\,\eV$.

For this reason, the resummation of \ir modes in \comet\hspace{-1pt}\vtwo is based on an alternative approach, the \dst algorithm, as originally proposed in \cite{HamHanLes2010} \citep[see e.g.][for applications of the same algorithm to independent codes]{ChuIvaPhi2020, IvaSimZal2020}. The algorithm involves performing a discrete sine transform of $\Plin$, followed by the removal of the bump associated with the \bao peak. This process is done by analyzing the second derivative of the sine transform, which identifies the location of the peak. Afterward, the gap created by removing the peak is filled using cubic-spline interpolation. Finally, the modified function is transformed back into Fourier space, resulting in a power spectrum that no longer contains \bao oscillations.

\begin{figure}
    \centering
    \includegraphics[width=\columnwidth]{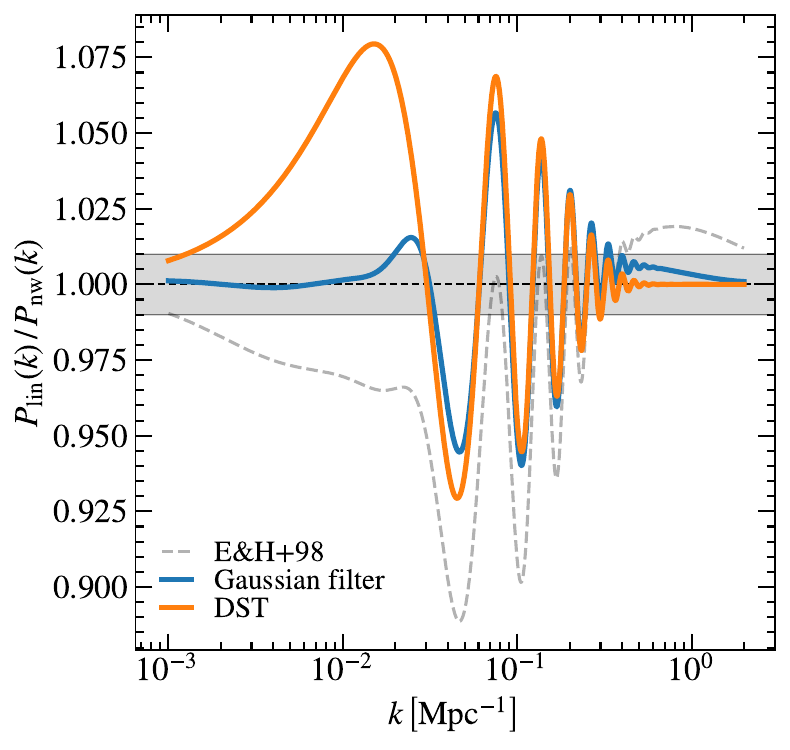}
    \caption{Ratio between the linear power spectrum $\Plin$ and the smooth power spectrum $\Pnw$ according to different modelling recipes, the Gaussian filtering technique and \dst, as specified in the legend. The reference cosmology assumed to generate the power spectra is a flat \lcdm one without massive neutrinos. The grey shaded band represents a 1$\%$ difference between the linear power spectrum and the no-wiggle template.}
    \label{fig:comp_nw}
\end{figure}

\begin{figure*}
    \centering
    \includegraphics[width=2\columnwidth]{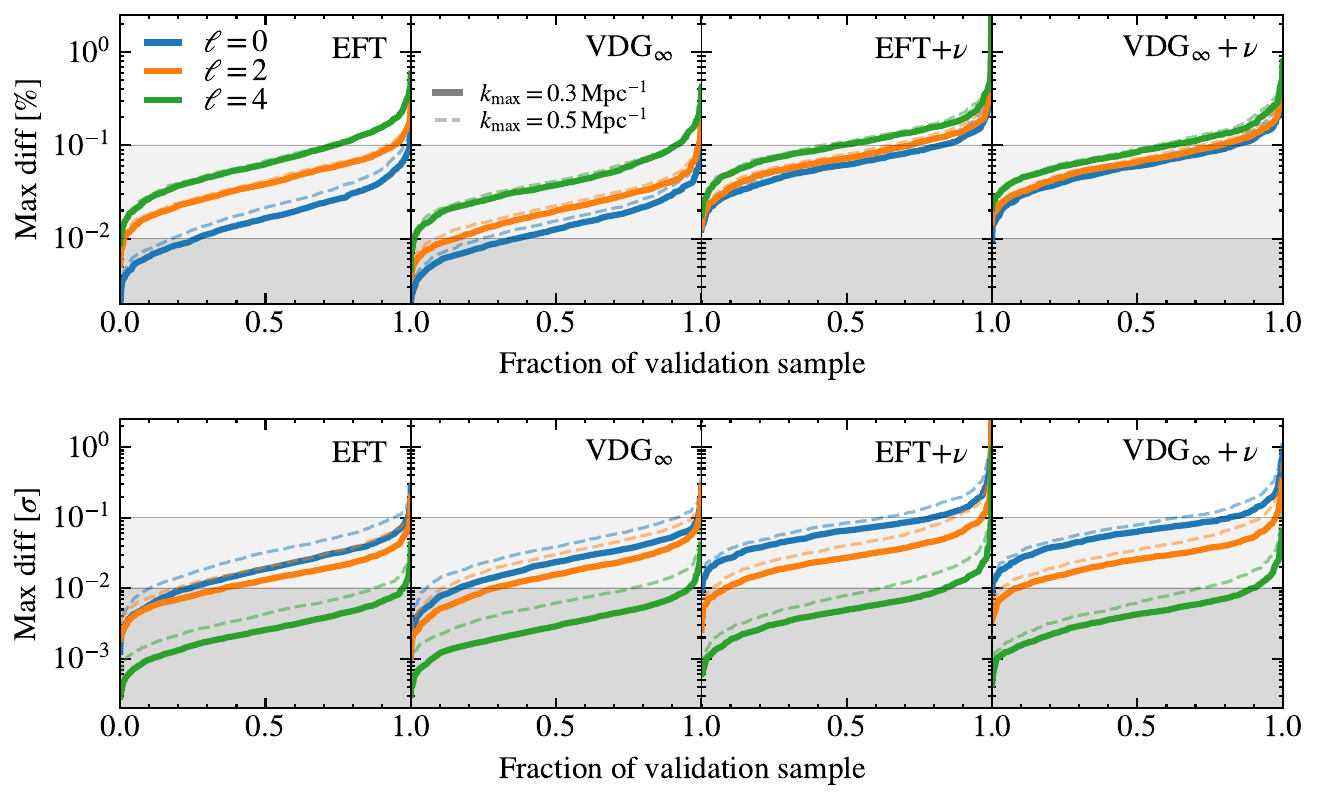}
    \caption{Cumulative distribution for the accuracy of the new set of emulators generated for \comet\hspace{-1pt}\vtwo. The x-axis represents the fraction of the validation sample, while the y-axis indicates the maximum difference between the emulator output and the validation set. This difference is measured either as the fractional percent deviation (top row, see Eq. \ref{eq:dist_percent}) or in terms of the Gaussian statistical noise from a sample with 10 times the volume of a redshift bin $z=(0.9,1.1)$ and an angular footprint of $15\,000\,{\rm sqdeg}$ (bottom row, see Eq. \ref{eq:dist_sigma}). The four columns correspond to four different emulators, for \eft and \vdg, each without and with massive neutrinos. The three different colors correspond to the distinct power spectrum multipoles, while the varying line styles represent the maximum mode $\kmax$ used for the validation.}
    \label{fig:accuracy_emulators}
\end{figure*}

Fig. \ref{fig:comp_nw} presents a comparison of the two methods for a single \lcdm cosmology without the inclusion of massive neutrinos. Even in this case, the aforementioned mismatch in the broadband shape of $\Peh$ and $\Plin$ is evident, but it is significantly reduced after application of the Gaussian filter. However, a residual deficiency in power is observed in $\Pnw$ on scales between $0.2\,\kMpc$ and $1\,\kMpc$, which can reach an amplitude of up to $1\%$. In contrast, the profile of the \dst $\Pnw$ aligns more closely with the actual mean position between the \bao wiggles, resulting in symmetric fluctuations of the wiggles around unity. In Sect.~\ref{sec:ir-resummation} we will study how these differences in the \ir-resummation schemes propagate to the outcomes of a likelihood analysis.

\section{Validation and performances}
\label{sec:ValPerf}

In this section, we finally present the validation of the new set of emulators, both in terms of their accuracy and their computational efficiency. Furthermore, we quantify the impact of the different \ir-resummation scheme adopted for \comet\hspace{-1pt}\vtwo in terms of the posterior distribution inferred from the analysis of multiple redshift bins with a combined volume consistent with the expected one from Stage-IV surveys, such as DESI and \Euclid. 

\subsection{Accuracy of the emulators}

To perform the validation of the new emulators, we constructed a dedicated validation set for each case. This validation set is generated from a distribution of \lcdm cosmological parameters, specifically $(\omb, \omc, \ns, \Mnu, h, \As)$. The shape parameters in this distribution have the same coverage as the training set, as detailed in Tab. \ref{tab:priors}, while the evolution parameters ($h$ on its own or in combination with $\As$ for the massless and massive neutrino case, respectively) are chosen from sufficiently large ranges. We then compute the corresponding power spectrum multipoles at $z=1$ to construct the final validation set. In this process, the bias, \rsd, and noise parameters are fixed to partially reproduce the clustering properties of the BOSS CMASS sample, as described in \cite{EggLeeSco2024}.

The accuracy estimation is performed using two different metrics. The first metric evaluates the distance $d(\rm{E,V})$ between the emulation and validation sets by calculating the maximum relative percent difference, defined as
\begin{equation}
    d_\%({\rm E\,, V}) = 100\,\max_i\left\{\frac{2\,\left|P_{\rm emu}(k_i)-P_{\rm val}(k_i)\right|}{\left[P_{\rm emu}(k_i)+P_{\rm val}(k_i)\right]}\right\}\,,
    \label{eq:dist_percent}
\end{equation}
where $P_{\rm emu}(k_i)$ and $P_{\rm val}(k_i)$ are the emulated and validation power spectrum multipoles at each wavenumber $k_i$, respectively. On the other side, to better understand the impact of the emulator inaccuracies in a real likelihood analysis, we define a new distance metric tied to a specific power spectrum multipole uncertainty $\sigma_P$, as
\begin{equation}
    d_\sigma({\rm E\,, V}) = \max_i\left\{\frac{\left|P_{\rm emu}(k_i)-P_{\rm val}(k_i)\right|}{\sigma_P(k_i)}\right\}\,.
    \label{eq:dist_sigma}
\end{equation}
For this exercise, similarly to the original validation of \comet, we consider the theoretical Gaussian uncertainty \citep{GriSanSal2016} from a volume enclosed in an angular footprint of $15\,000\,{\rm sqdeg}$ and redshift bins of width $\Delta z=0.2$. To conduct more stringent validation tests, the volume is further multiplied by a factor of ten, enhancing the sensitivity of the analysis.

The results of the accuracy tests are shown in Fig. \ref{fig:accuracy_emulators}, where we present both distance metrics for the full set of emulators that we have trained. To ensure completeness, two sets of curves are shown, one calculated over the entire range of wavenumbers used to train the emulators ($\kmax=0.5\,\kMpc$) and another limited to a more realistic scale cut, considering a more sensible range of validity of these perturbative models ($\kmax=0.3\,\kMpc$).\footnote{As pointed out in \cite{EggLeeSco2024}, the \vdg model can be used to robustly describe the shape of the galaxy power spectrum multipoles up to this value of $\kmax$, when considering \hod mocks based on the BOSS CMASS and LOWZ galaxy samples.}

In both cases, we observe that the accuracy remains consistently better than $0.3\%$ for all configurations. The relative accuracy tends to be higher when considering lower-order multipoles and scenarios without massive neutrinos. This trend is reversed when using the second distance metric, where the tighter error bars on the monopoles cause the emulated quantities to appear more discrepant from the validation set compared to higher-order multipoles. Even when including massive neutrinos, the majority of the sample remains consistent with the validation set, achieving an accuracy that is better than $0.3\sigma$.

\begin{figure}
    \centering
    \includegraphics[width=\columnwidth]{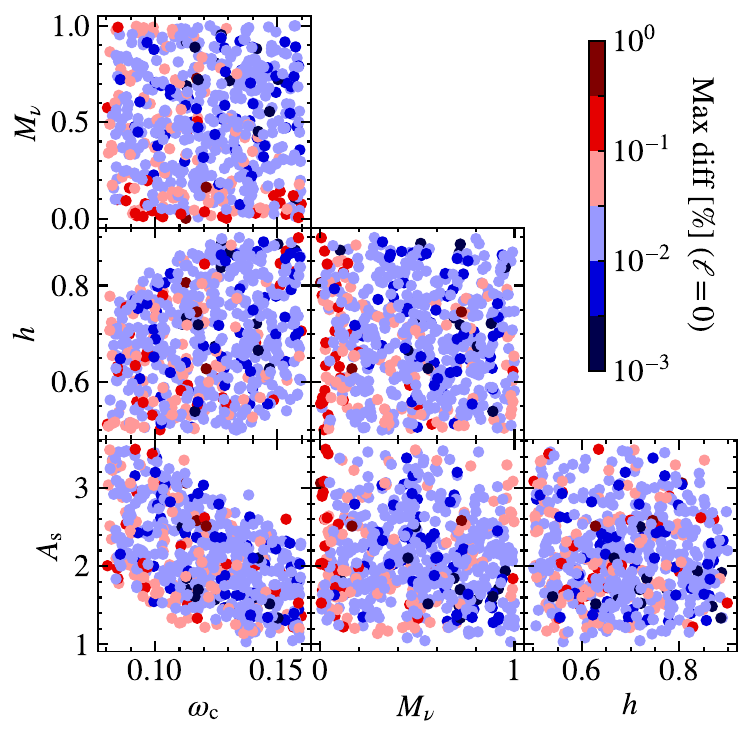}
    \caption{Distribution of the relative percentage accuracy for the monopole of the EFT model, based on the corresponding validation set. The plot displays the parameters $(\omc, \Mnu, h, \As)$, while we marginalise over the remaining parameters $(\omb, \ns)$. Data points are color-coded according to the maximum relative difference.}
    \label{fig:accuracy_v2_triangle}
\end{figure}

In addition to assessing the overall accuracy of the new emulators, we are particularly interested in verifying that the extended recipe for evolution mapping in the presence of massive neutrinos performs equally well regardless of the total neutrino mass $\Mnu$. Fig. \ref{fig:accuracy_v2_triangle} illustrates the accuracy distribution estimated from the validation set as a function of the position in cosmological parameter space. The full parameter space is marginalized over $(\omb, \ns)$, so we only display the accuracy distribution for the remaining parameters. Since the validation set is constrained to samples where $\sigtwelvestar$ and $f$ deviate by no more than 20\% from the corresponding \Planck predictions at the same redshift, the multidimensional distribution does not encompass the entire parameter space, as evident in, e.g., the $\omc$~-~$\As$ panel. As the figure shows, we do not observe any significant dependence of the emulator accuracy on neutrino mass. We only note a slight increase in the maximum differences (on average at the level of $0.1\%$) for low $\Mnu$ values, likely reflecting the proximity to the edge of the parameter space for which the emulators have been trained.

\subsection{Computational performance}

\begin{figure}
    \centering
    \includegraphics[width=\columnwidth]{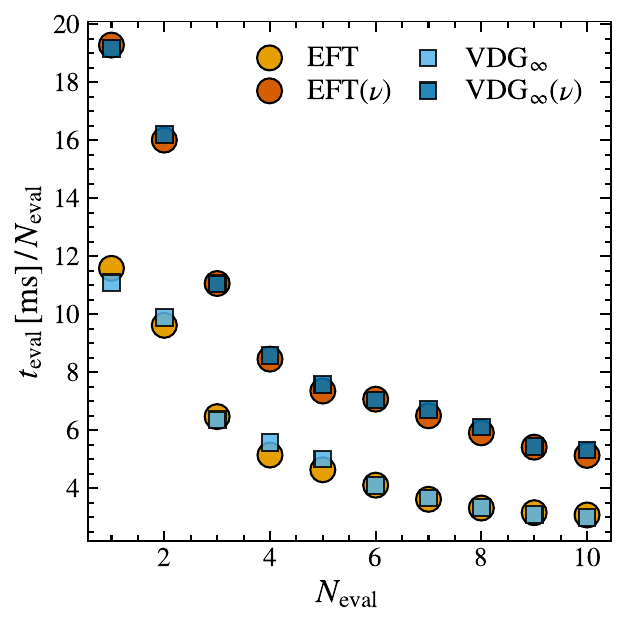}
    \caption{Evaluation time required to compute the three power spectrum multipoles ($\ell=(0,2,4$) across different emulator models, indicated in the legend. The time per evaluation is plotted against the number of batch evaluations, illustrating the efficiency gain when generating multiple samples simultaneously compared to running individual emulator calls for each sample separately.}
    \label{fig:evaluation_time}
\end{figure}

In terms of computational efficiency, the new set of emulators aligns closely with the performance reported in \cite{EggCamPez2022}, even with the increased demands of the \comet\hspace{-1pt}\vtwo setup, which includes an extended $k$ range and a broader parameter space. Fig. \ref{fig:evaluation_time} illustrates the scaling of the total evaluation time for predicting the galaxy power spectrum multipoles (considering 100 $k$ bins) with respect to the number of batch evaluations introduced in Sect. \ref{sec:batch_evaluation}. A single model evaluation on a single-core CPU takes approximately $11\,{\rm ms}$ for both the \eft and \vdg models without massive neutrinos. On the other hand, emulators with massive neutrinos require more time per evaluation due to the expanded training set and parameter space, with a single evaluation taking around $19\,{\rm ms}$. As discussed in Sect. \ref{sec:batch_evaluation}, performing $N_{\rm eval}$ evaluations simultaneously reduces the overall evaluation time compared to executing $N_{\rm eval}$ individual calls. When considering more than two simultaneous evaluations, the slope of the $t_{\rm eval}(N_{\rm eval})$ relation becomes shallower, leading to the gains in computation time already mentioned in Sect. \ref{sec:batch_evaluation}.

Overall, we find that replacing the direct computation of the real non-linear model with a \comet evaluation improves computational speed by approximately three orders of magnitude. This acceleration is crucial for enabling efficient statistical inference of cosmological parameters in a likelihood analysis. Additionally, using evolution mapping to account for different evolution parameters reduces the dimensionality of the parameter space. In turn, this improves the accuracy of the emulator at a fixed evaluation time.

\subsection{Impact of different \ir-resummation schemes}
\label{sec:ir-resummation}

In this section, we examine the impact of changing methods for isolating the no-wiggle power spectrum $\Pnw$, which is required for the implementation of \ir-resummation. As noted in Sect. \ref{sec:ir-resum}, \comet\hspace{-1pt}\vtwo replaces the original Gaussian filtering technique applied to the featureless template of \cite{EisHu1999} with the \dst algorithm. In Fig. \ref{fig:comp_nw} we already compared these methods at the level of $\Pnw$, whereas here we are interested in determining how the residual differences propagate to the galaxy power spectrum multipoles and, ultimately, to the posterior distribution of cosmological parameters. 

\begin{figure}
    \centering
    \includegraphics[width=\columnwidth]{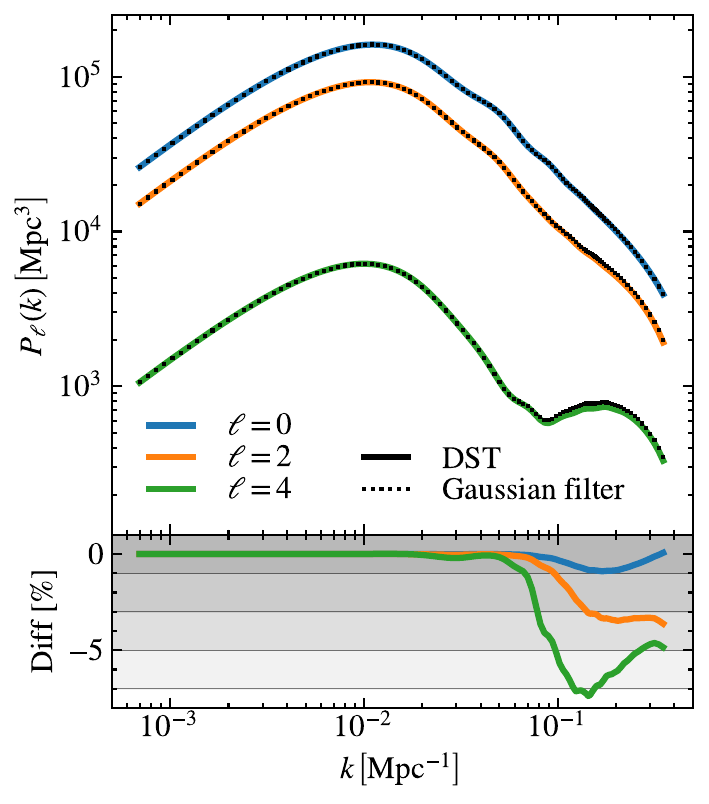}
    \caption{\emph{Top:} Comparison of the power spectrum multipoles within the EFT framework, using either the Gaussian filter or the DST methods to account for \ir-resummation effects. This analysis uses bias and \rsd parameters derived from fits to synthetic mocks of H$\alpha$ galaxies, and assumes a power spectrum amplitude of $\sigtwelve = 0.7$. \emph{Bottom}: Percentage differences between the two approaches, with grey shaded areas highlighting the 1$\%$, 3$\%$, 5$\%$, and 7$\%$ intervals.}
    \label{fig:different_IR}
\end{figure}

The comparison in terms of the final model for the $\Pell(k)$ is illustrated in Fig. \ref{fig:different_IR}, where we use a fixed set of cosmological and nuisance parameters to plot the power spectrum Legendre multipoles, employing both the Gaussian smoothing technique and the \dst framework. The relative differences in the bottom panel reveal that discrepancies emerge on mildly non-linear scales ($k\gtrsim0.1\,\kMpc$) and are more pronounced for higher-order multipoles. The residuals range from $1\%$ for the monopole $P_0$ to over $5\%$ for the hexadecapole $P_4$. This level of discrepancy has also been reported in previous studies \citep[e.g.][]{LinMorRad2024}, raising the question of whether such differences could ultimately result in inconsistent cosmological constraints in a realistic likelihood analysis.

\begin{figure*}
    \includegraphics[width=2\columnwidth]{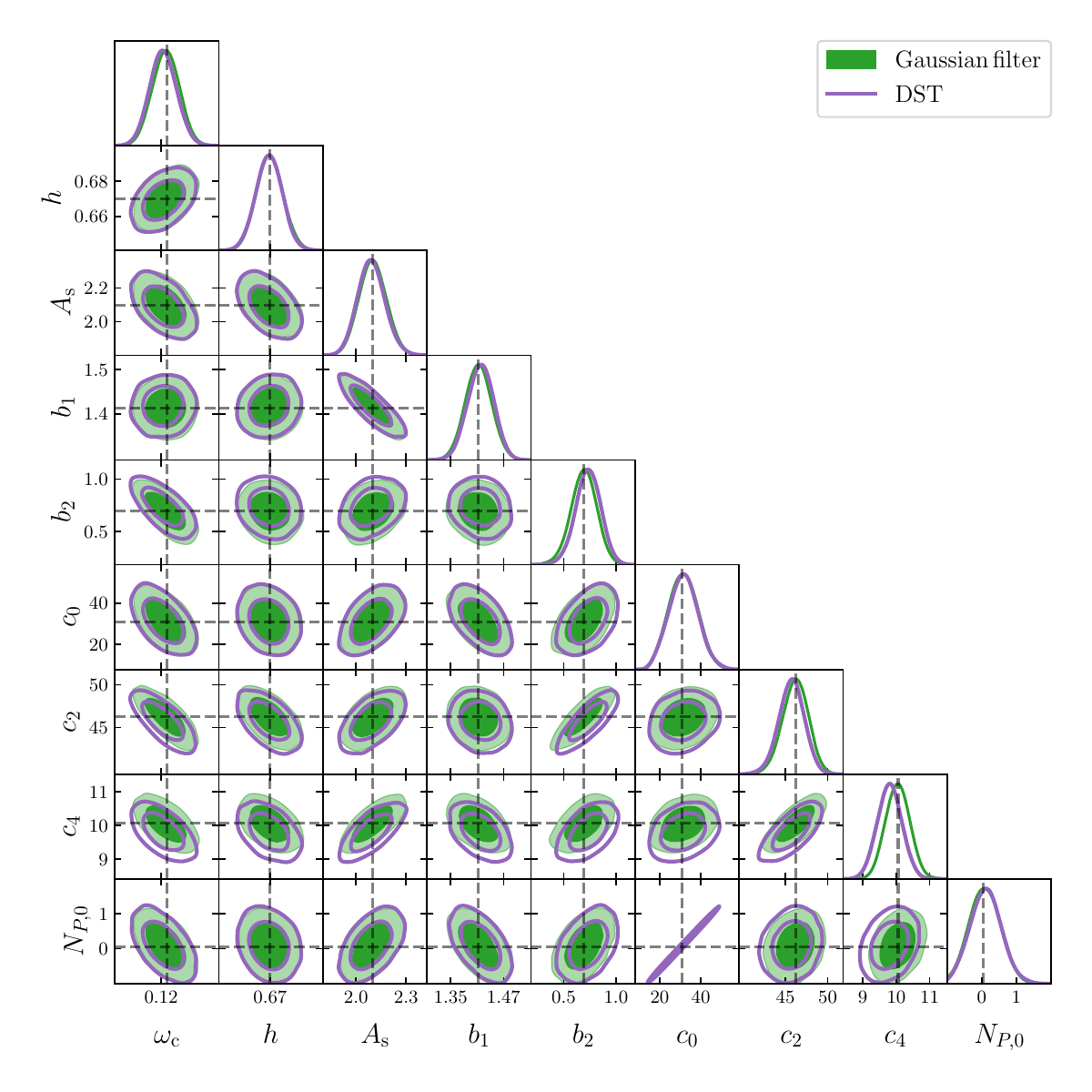}
    \caption{Marginalised posterior distribution of the cosmological parameters $(\omc, h, \As)$ and the nuisance parameters $(\bone, \btwo, \czero, \ctwo, \cfour, \npzero)$ for the \eft modelling framework. The synthetic data vectors are generated at $z=1$ using the Gaussian filter technique, applied to the smooth power spectrum from \cite{EisHu1999}. The two distinct marginalised contours represent cases where the theoretical model used in the likelihood is either the Gaussian filter or the \dst method. In both scenarios, a Gaussian covariance matrix is assumed, derived from a redshift shell with width $\Delta z=0.2$ and an angular footprint of $15\,000\,{\rm sqdeg}$, with the fit extending up to $\kmax=0.3\,\kMpch$.}
    \label{fig:posterior_different_IR}
\end{figure*}

To address this issue, we generate a synthetic data vector consisting of the monopole, quadrupole, and hexadecapole at $z=1$ obtained from the \eft model with the Gaussian filter approach. We then analyze this synthetic data vector using the \eft model combined with either the Gaussian filter or the \dst approach. For this test, we adopt a Gaussian covariance matrix based on the predictions of \cite{GriSanSal2016} (a conservative choice for this test) along with a volume and sample number density that is representative for a Stage-IV galaxy survey. Specifically, we use a volume corresponding to a single redshift bin of size $\Delta_z=0.2$ over a sky area of $15000\,\sqdeg$, and a number density based on current forecasts for the spectroscopic H$\alpha$ sample \citep{PozHirGea2016, CasFosSta2024, PezMorZen2024} that will be targeted by \Euclid \citep{EUCLID2011, MelAbdAce2024}. In the likelihood analysis we sample three cosmological parameters ($\omc,\,h
,\,\As$) and six nuisance parameters ($\bone,\,\btwo,\,\czero,\,\ctwo,\,\cfour,\,\npzero$), and their posteriors are shown in Fig. \ref{fig:posterior_different_IR}. 

Despite the residual differences between the two \ir-resummation methods, the figure demonstrates that the impact on cosmological parameters is minimal, with only a slight but negligible shift in the $\omc$ posterior. In contrast, more significant shifts are observed for some of the nuisance parameters, particularly the counterterms $\ctwo$ and $\cfour$. This is expected, given that they contribute almost exclusively to the quadrupole $P_2$ and the hexadecapole $P_4$, respectively, which -- as shown in Fig. \ref{fig:different_IR} -- are the multipoles most affected by the different \ir-resummation methods.\footnote{We note that this statement may not hold when considering a parametrization of the \rsd counterterms different from the one used in \comet. In our approach, each of the three leading-order terms primarily influences the amplitude of the corresponding power spectrum multipole. However, with an alternative parametrization -- such as one using powers of $\mu$ instead of the Legendre polynomials ${\cal L}_\ell(\mu)$ -- some of the observed discrepancies in $\ctwo$ and $\cfour$ could potentially leak into $\czero$.} Other parameters, such as $\btwo$, also exhibit slight deviations from the fiducial values used to generate the data vectors, although these deviations remain within $1\sigma$ of the expected values. This indicates that the residual differences introduced by the different \ir-resummation schemes can be effectively absorbed by the nuisance parameters, without significantly impacting the cosmological parameters.

\section{Conclusions}
\label{sec:conclusions}





This work extends the evolution mapping framework described in \cite{SanRuiJar2022} to cosmological models that include massive neutrinos. In its original formulation, evolution mapping describes the evolution of the matter power spectrum only in terms of the shape parameters and a single amplitude parameter, $\sigtwelve$. This can then be used as a new time variable to map different cosmologies with the same shape parameters into each other. This approach breaks down in the presence of massive neutrinos due to the suppression of power on scales below their typical free-streaming scale, since the strength of the damping is not uniquely determined in terms of the total neutrino mass $\Mnu$ and $\sigtwelve$. To restore the mapping across different cosmologies with massive neutrinos, we developed a more general formulation which also accounts for the mapping of the neutrino suppression factor $S(k)$. This is achieved in terms of the combination $\sigtwelvestar/\sqrt{\As}$, where $\sigtwelvestar$ is evaluated for a cosmology with the same total matter density of the one under consideration but composed solely of cold dark matter and baryons.


With these minor adjustments, it remains possible to achieve an almost exact mapping (to within $\sim 0.01\%$ accuracy) between cosmologies with different evolution parameters -- such as expansion rate, curvature, dark energy, and redshift -- and for a broad range of total neutrino masses (at least up to $\Mnu=1\,\eV$). We confirmed that this holds true both when considering the power spectrum of cold dark matter and baryons, as well as the total matter density field, which also includes fluctuations in the neutrino density field. Furthermore, this approach is equally effective when varying the number of massive neutrino species, as we verified by explicitly testing configurations with $\Nnu=1$ and $\Nnu=3$.

As a follow-up to this result, we applied the extended evolution mapping framework to develop a new set of emulators of the galaxy power spectrum for cosmologies including massive neutrinos within the public \comet package \cite{EggCamPez2022}. This involved expanding the input parameter space to include both $\Mnu$ and $\As$ with sufficiently broad ranges, in order to encompass a wide variety of cosmological models. Specifically, we sampled $\Mnu$ in the range $[0, 1] \,\eV$ and $\As$ in the range $[1.0, 3.5] \cdot 10^{-9}$. Additionally, we extended the prior ranges for the shape parameters already present in \comet, increasing them by factors of 1.65, 1.15, and 1.45 for $\omb$, $\omc$, and $\ns$, respectively. Given that the accuracy and computational performance of the emulators are affected by the two extra dimension of the parameter space, we decided to retain a newly trained version of the emulators without massive neutrinos as separate tools. Users can select between the standard or massive neutrino versions by setting a dedicated flag, depending on the requirements of their analysis.

The newly created \comet emulator has been trained under the assumption of one massive and two massless neutrino species, with a total effective number of species fixed at $\Neff=3.044$. We note that modifications to either the number of massive species or the value of 
$\Neff$ would impact the shape of the matter power spectrum and thus require a new dedicated training set. Future releases of \comet will explore these directions to support a broader range of cosmological scenarios.

Consistently with the latest version of the software, \comet\hspace{-1pt}\vtwo allows users to obtain predictions using either the \eft or \vdg model (plus the real-space perturbative model), ensuring flexibility in the choice of theoretical framework. Similarly to the choice of including massive neutrinos or not, users can specify their preferred \rsd model through a flag when initially creating an instance of the emulator. Beyond this selection, the code operates almost identically for both models.

Apart from the expanded parameter space, the new version of \comet introduces several additional features compared to its predecessor. For example, it offers an extended support for the wavemode $k$, as the new emulators have been trained up to $\kmax = 0.5\, \kMpc$, an increase from the original limit of $\kmax = 0.35\, \kMpc$. Most notably, the new version allows for simultaneous predictions of multiple Legendre multipoles with a single call to the emulator. This improvement, coupled with the transition to a different Gaussian process software, has reduced the total evaluation time by approximately $50\%$. Such efficiency gains can offer significant advantages, particularly when combining galaxy clustering constraints from multiple redshift bins or from different data sets into a single posterior distribution. The process of sampling the posterior distribution over a given parameter space can be further accelerated through the implementation of analytical marginalization. This approach reduces the number of sampled bias, \rsd, and noise parameters to only 3–4 per spectroscopic bin, depending on the chosen \rsd model. This feature can be activated by the user when utilizing the built-in routine to calculate the $\chi^2$ statistic between the theoretical model and user-provided data vectors.

Finally, the new sets of emulators include a different \ir-resummation scheme, compared to the original version of \comet. This transition was motivated by the poor performances of the Gaussian filtering technique applied to the featureless transfer function of \cite{EisHu1999} with cosmologies including massive neutrinos. Specifically, using this approach, residuals in the wiggle-no-wiggle split are observed at mildly non-linear scales, potentially affecting the construction of the final galaxy power spectrum model. The new technique introduced in \comet\hspace{-1pt}\vtwo is based on a discrete sine transform, as also used in other public software packages \citep[i.e.,][]{ChuIvaPhi2020}. This approach successfully eliminates residuals, even for cosmologies that include massive neutrinos. We quantify the impact of different \ir-resummation schemes on the final galaxy power spectrum multipoles for a neutrino-free cosmology and find that higher-order multipoles are more sensitive to the chosen method, with residuals ranging from ${\cal O}(1\%)$ in the monopole up to ${\cal O}(10\%)$ in the hexadecapole. Despite these differences, we show that the inference of cosmological parameters is only minimally affected, with noticeable deviations confined to specific \rsd parameters, such as the \eft counterterms. Since these parameters are typically marginalized in Bayesian analyses, we conclude that switching between \ir-resummation schemes can be considered safe, depending on user preference.

This paper is accompanied by the public release of \comet\hspace{-1pt}\vtwo, which incorporates all the new features outlined in the main body of this article and summarized in the previous sections. The updated code can be installed via \texttt{PyPI}\,\footnote{\url{https://pypi.org/project/comet-emu/}} or by cloning the corresponding \texttt{git} repository.\footnote{\url{https://gitlab.com/aegge/comet-emu}}

\section*{Acknowledgements}

We acknowledge useful discussions with Matteo Esposito, Daniel Farrow, Martha Lippich, and Agne Semenaite.
AP acknowledges support from the INAF AstroFit fellowship (Iniziative di Ricerca Fondamentale 2021-2025). AE is supported at the Argelander Institut für Astronomie by an Argelander Fellowship. GG, BC, MC, and GP acknowledge support from the Spanish Ministerio de Ciencia, Innovación y Universidades, project PID2021-128989NB. AGS is supported by the Deutsche Forschungsgemeinschaft (DFG, German Research Foundation) under Germany´s Excellence Strategy – EXC 2094 – 390783311. NL was supported by the James Arthur Graduate Associate Fellowship from the Center for Cosmology and Particle Physics at New York University and the Horizon Fellowship from Johns Hopkins University. Our research made use of \texttt{matplotlib}, a \python \ library for publication quality graphics \cite{Hun2007}.

\appendix

\section{Training set boundaries for correlated parameters}
\label{sec:app.boundaries}

To reduce the number of samples needed for training the emulator, we exploit the fact that some of the parameters exhibit stronger correlations, specifically between $f$ and $\sigtwelvestar$, as well as $A_s$ and $\omc$. Based on these correlations, we can define boundaries, $f_{\rm max}(\sigtwelvestar)$ and $f_{\rm min}(\sigtwelvestar)$, or $A_{\rm s,max}(\omctilde)$ and $A_{\rm s,min}(\omctilde)$, which indicate regions of the parameter space that are unlikely to be sampled in any realistic application of the emulator, allowing us to safely exclude them from the construction of our training dataset. Following the procedure outlined in Sect.~\ref{sec:parameter-ranges} and combining with the strict limits $0.48 \leq f \leq 1.05$ and $\As \leq 3.5\,\times\,10^{-9}$, we determine the boundaries:
\begin{widetext}
\begin{align}
  f_{\rm max}(\sigtwelvestar) &\approx \left\{
    \begin{array}{ll}
        1.05\,,  &  \text{for} \quad \sigma_{12,\star} \leq 0.44 \\
        -0.73\,\sigma_{12,\star}^2 + 0.19\,\sigma_{12,\star} + 1.11\,, & \text{otherwise}
    \end{array}\right.\,, \\
  f_{\rm min}(\sigma_{12,\star}) &\approx \left\{
    \begin{array}{ll}
        -1.69\,\sigma_{12,\star}^2 + 0.59\,\sigma_{12,\star} + 0.82\,, &  \text{for} \quad \sigma_{12,\star} \leq 0.65 \\
        0.48\,, & \text{otherwise}
    \end{array}\right.\,,
\end{align}
and
\begin{align}
  A_{\rm s,max}(\tilde{\omega}_c) &\approx 10^{-9}\,\left\{
  \begin{array}{ll}
    3.5\,, &  \text{for} \quad \tilde{\omega}_c \leq 0.85 \\
    2.1\,\omctilde^2 - 5.5\,\omctilde + 4.9\,, & \text{otherwise}
  \end{array}\right.\,, \\
  A_{\rm s,max}(\omctilde) &\approx 10^{-9}\,\left(0.45\,\omctilde^2 - 1.6\,\omctilde + 1.8\right)\,,
\end{align}
where $\omctilde \equiv \omc/0.1198$.
\end{widetext}

\bibliography{main}

\end{document}